\documentclass[10pt]{article}
\usepackage{graphicx}
\usepackage[T1]{fontenc}
\usepackage[utf8]{inputenc}
\usepackage{amssymb}
\usepackage{amsfonts}
\usepackage{dsfont}
\usepackage{mathtools}
\usepackage{amsthm}
\usepackage{amsmath}
\usepackage{textcomp}
\usepackage{eurosym}

\usepackage{bigints}
\usepackage{geometry}
\newtheorem{rem}{Remark}
\newtheorem{assumption}{Assumption}

\begin{document}
\title{Algorithmic market making for options\footnote{Bastien Baldacci gratefully acknowledges the support of the ERC Grant 679836 Staqamof. Olivier Gu\'eant thanks the Research Initiative ``Modélisation des marchés actions, obligations et dérivés'' financed by HSBC France under the aegis of the Europlace Institute of Finance for their support regarding an early version of the paper (entitled ``Algorithmic market making: the case of equity derivatives''). The authors would like to thank Lorenzo Bergomi (Société Générale), Bruno Bouchard (Université Paris-Dauphine), Rama Cont (Oxford University), Renaud Delloye (HSBC), Thomas de Garidel (HSBC), Nicolas Grandchamp des Raux~(HSBC), Iuliia Manziuk (Université Paris 1 Panthéon-Sorbonne), Ben Nasatyr (Citi), Jiang Pu (Institut Europlace de Finance), and Mathieu Rosenbaum (Ecole Polytechnique) for the discussions they had on the topic. The readers should nevertheless be aware that the views, thoughts, and opinions expressed in the text belong solely to the authors.
}}
\author{Bastien {\sc Baldacci}\footnote{\'Ecole Polytechnique, CMAP, Route de Saclay, 91128 Palaiseau Cedex, France,  bastien.baldacci@polytechnique.edu.} \and Philippe {\sc Bergault}\footnote{Université Paris 1 Panthéon-Sorbonne, Centre d'Economie de la Sorbonne, 106 Boulevard de l'Hôpital, 75642 Paris Cedex~13, France, philippe.bergault@etu.univ-paris1.fr.} \and Olivier {\sc Guéant}\footnote{Université Paris 1 Panthéon-Sorbonne, Centre d'Economie de la Sorbonne, 106 Boulevard de l'Hôpital, 75642 Paris Cedex~13, France, olivier.gueant@univ-paris1.fr. \emph{Corresponding author.}}}
\date{}
\maketitle

\begin{abstract}
\noindent In this article, we tackle the problem of a market maker in charge of a book of options on a single liquid underlying asset. By using an approximation of the portfolio in terms of its vega, we show that the seemingly high-dimensional stochastic optimal control problem of an option market maker is in fact tractable. More precisely, when volatility is modeled using a classical stochastic volatility model -- e.g. the Heston model -- the problem faced by an option market maker is characterized by a low-dimensional functional equation that can be solved numerically using a Euler scheme along with interpolation techniques, even for large portfolios. In order to illustrate our findings, numerical examples are provided.
\end{abstract}

\vspace{5mm}


\setlength\parindent{0pt}

\section{Introduction}

The electronification of financial markets started in the seventies with stock exchanges and now affects each and every asset class. For asset classes that are usually traded in a centralized way (stocks, futures, etc.), exchanges and other all-to-all trading platforms -- based or not based on limit order books -- are now fully automated. For assets that are still traded over the counter (OTC), electronification occurs through the introduction of new platforms, for instance single- and multi-dealer-to-client platforms.\\

This electronification is associated with a trend towards the automation of the trading process for many players of the financial industry: brokers, banks, but also systematic asset managers who often develop their own execution algorithms. For assets traded in centralized markets, trading automation is nowadays massive. For instance, in the cash equity world, a vast majority of the execution is now carried out using algorithms. For assets traded in dealer markets, the automation of the market making process has been at the agenda for a few years and more and more banks are developing market making algorithms for various asset classes (currencies, bonds, etc.).\\

In the academic literature, many market making models have been proposed since the eighties. In the early literature on market making, the two main references are the paper of Ho and Stoll~\cite{ho1981optimal} and the paper of Grossman and Miller~\cite{grossman1988liquidity}. Ho and Stoll introduced indeed a very relevant framework to tackle the main problem faced by market makers: inventory management. Grossman and Miller, who were more interested in capturing the essence of liquidity, proposed a very simple model with 3 periods that encompassed both market makers and final customers, enabled to understand what happens at equilibrium, and contributed to the important literature on the price formation process. If the latter paper belongs to a strand of literature that is extremely important to go beyond the simple Walrasian view of markets, it is of little help to build market making algorithms. The former paper however, after more than 25 years, has paved the way to a recent mathematical literature on algorithmic market making.\\

The seminal reference of the new literature on market making is the paper of Avellaneda and Stoikov~\cite{avellaneda2008high} who revived the dynamic approach proposed by Ho and Stoll. They indeed showed how the quoting and inventory management problems of market makers could be addressed using the tools of stochastic optimal control. Since then, many models have been proposed, most of them to tackle the same problem of single-asset market making as that of Avellaneda and Stoikov. For instance, Guéant, Lehalle, and Fernandez-Tapia provided in~\cite{gueant2013dealing} a rigourous analysis of the Avellaneda-Stoikov stochastic optimal control problem and proved that the problem could be simplified into a system of linear ordinary differential equations (ODE) in the case of exponential intensity functions. Cartea, Jaimungal, and coauthors contributed a lot to the literature and added many features to the initial models:  alpha signals, ambiguity aversion, etc. (see~\cite{cartea2017algorithmic,cartea2015algorithmic,cartea2014buy}). They also considered a different objective function: a risk-adjusted expectation instead of a Von Neumann-Morgenstern expected utility.\\

The models proposed in the above papers all share the same characteristics: (i) they are agnostic with respect to the market structure but are in fact more adapted to OTC markets,\footnote{The models are also suited for assets traded in limit order books if the bid-ask spread to tick size ratio is large.} (ii) they only deal with single-asset market making, and (iii) they do not deal with the market making of options.\\

In fact, models have been specifically developed by Guilbaud and Pham (see~\cite{guilbaud2013optimal,guilbaud2015optimal}) for assets traded through limit order books (e.g. most stocks) and for assets traded on platforms with a pro-rata microstructure (e.g. some currency pairs). Interestingly, these models enable the use of aggressive orders by market makers, which is -- surprisingly -- a standard behavior on equity markets (see~\cite{pamela}).\\

As far as multi-asset market making is concerned, models have been developed recently to account for the correlation structure between asset price changes. Guéant extended to a multi-asset framework both models \emph{à la} Avellaneda-Stoikov and models \emph{à la} Cartea-Jaimungal (see~\cite{gueant2016financial},~\cite{gueant2017optimal}, and~\cite{gueant2015general}) and showed that the problem boils down, for general intensity functions, to solving a system of (\emph{a~priori} nonlinear) ODEs. The associated question of the numerical methods to approximate the solution of the equations characterizing the optimal quotes of a multi-asset market maker is addressed in~\cite{bergault2019optimal} using a factorial approach and in~\cite{gueant2019deep} using reinforcement learning, both with applications to corporate bond markets.\\

Finally, as far as asset classes are concerned, there have been few attempts to address market making problems outside of the cash world. Market making models for derivative contracts are indeed intrinsically more complicated because they must account for the strategies on both the market for the underlying asset and the market for the derivatives, and usually for numerous contracts (e.g. options for lots of strikes and maturities). Option market making is only addressed in a paper by El Aoud and Abergel (see~\cite{el2015stochastic}) and in a paper by Stoikov and Sa{\u{g}}lam~\cite{stoikov2009option}. In the former, the authors consider a single-option market driven by a stochastic volatility model and assume that the position is always $\Delta$-hedged. They provide optimal bid and ask quotes for the option and focus on the risk of model misspecification. In the latter, the authors consider three different settings, but all with only one option: (i) a market maker in a complete market where continuous trading in the perfectly liquid underlying stock is allowed, (ii) a market maker who may not trade continuously in the underlying stock, but rather sets bid and ask quotes in the option and the stock, and (iii) a market maker in an incomplete market with residual risks due to stochastic volatility and overnight jumps.\\

In this paper, we consider the case of a market maker in charge of a book of options whose prices are driven by a stochastic volatility model. We assume that trading in continuous time can be carried out in the underlying asset so that the residual risk is only that of the vega associated with the inventory. Using a constant-vega approximation, we show that the problem of an option market maker boils down to solving a low-dimensional functional equation of the Hamilton-Jacobi-Bellman type that can be tackled numerically using a simple Euler scheme along with interpolation techniques. In particular, in spite of the large number of assets, the market making problem is tractable. \\

In Section 1, we describe the model and present the optimization problem of the option market maker. In Section 2, we show how that problem can be simplified under the constant-vega approximation. In particular, we show that solving the high-dimensional stochastic optimal control problem of the market maker boils down to solving a low-dimensional functional equation. In Section 3, we consider the example of a book of options with several strikes and maturities and provide numerical results obtained through interpolation techniques and an explicit Euler scheme.\\

\section{Description of the problem}

We consider a probability space $(\Omega,\mathcal{F},\mathbb{P}\big)$ with a filtration $(\mathcal{F}_{t})_{t\in \mathbb{R}_{+}}$ satisfying the usual conditions. Throughout the paper, we assume that all stochastic processes are defined on $\big(\Omega,\mathcal{F},(\mathcal{F}_t)_{t\in \mathbb{R}_{+}},\mathbb{P}\big)$.

\subsection{The market}

\medskip
We consider an asset whose price dynamics is described by a one-factor stochastic volatility model of the form
\begin{align*}
\left\{
    \begin{array}{ll}
        dS_{t} = \mu S_{t}dt +\sqrt{\nu_{t}}S_{t}dW_{t}^{S}  \\
        d\nu_{t}=a_{\mathbb{P}}(t,\nu_t)dt+\xi\sqrt{\nu_{t}}dW_{t}^{\nu},
    \end{array}
\right.
\end{align*}
where $\mu \in \mathbb{R},\xi \in \mathbb{R}_+^{*}$, $(W_{t}^{S},W_{t}^{\nu})_{t\in \mathbb{R}_{+}}$ is a couple of Brownian motions with quadratic covariation given by $\rho=\frac{d\langle W^{S},W^{\nu}\rangle}{dt}  \in (-1,1)$, and
$a_{\mathbb{P}}$ is such that the processes are well defined (in particular, we assume that the process $(\nu_t)_{t\in \mathbb{R}_{+}}$ stays positive almost surely).\\

\begin{rem}
  A classical example for the function $a_{\mathbb{P}}$ is that of the Heston model (see~\cite{heston1993closed}), i.e. $a_{\mathbb{P}}: (t,\nu) \mapsto \kappa_{\mathbb{P}}(\theta_{\mathbb{P}} - \nu)$ where $\kappa_{\mathbb{P}},\theta_{\mathbb{P}} \in \mathbb R^*_+$ satisfy the Feller condition $ 2 \kappa_{\mathbb{P}}\theta_{\mathbb{P}} > \xi^2$.\\
\end{rem}

\begin{rem}
For the sake of simplicity, we consider throughout this paper a one-factor model where the instantaneous variance is the main variable of interest. Similar results could be obtained with a one-factor model focused on forward variances, such as the classical one-factor Bergomi model (see \cite{bergomi2005smile, bergomi2015stochastic}). Moreover, it is noteworthy that our approach can easily be extended to two-factor stochastic volatility models such as the celebrated two-factor Bergomi model (see \cite{bergomi2005smile, bergomi2015stochastic}), up to an increase -- by $1$ -- of the dimension of the equation to solve.\\
\end{rem}

Assuming interest rates are equal to $0$, we introduce an equivalent risk-neutral/pricing probability measure\footnote{For references, see for instance~\cite{gatheral2011volatility}.} $\mathbb{Q}$ under which the price and volatility processes become
\begin{align*}
\left\{
    \begin{array}{ll}
        dS_{t} = \sqrt{\nu_{t}}S_{t}d\widehat{W}_{t}^{S}  \\
        d\nu_{t}=a_{\mathbb{Q}}(t,\nu_t)dt+\xi\sqrt{\nu_{t}}d\widehat{W}_{t}^{\nu},
    \end{array}
\right.
\end{align*}
where $(\widehat{W}_{t}^{S},\widehat{W}_{t}^{\nu})_{t\in \mathbb{R}_{+}}$ is another couple of Brownian motions, this time under $\mathbb{Q}$, with quadratic covariation given by $\rho=\frac{d\langle \widehat{W}^{S},\widehat{W}^{\nu}\rangle}{dt}  \in (-1,1)$, and where $a_{\mathbb{Q}}$ is such that the processes are well defined.\\

We consider $N\geq 1$ European options written on the above asset (hereafter, the underlying asset). For each $i \in \left\{1, \dots, N \right\}$, the maturity date of the $i$-th option is denoted by $T^{i}$ and we denote by $(\mathcal{O}_{t}^{i})_{t\in [0,T^i]}$ the price process associated with the $i$-th option.\\

\begin{rem}
  In applications, the options under consideration will always be call and/or put options. However, our setting enables to consider any European payoff.\\
\end{rem}

In the above one-factor model, we know that for all $i \in \left\{1, \dots, N \right\}$, and all $t\in [0,T^{i}]$,  $\mathcal{O}_{t}^{i}=O^{i}(t,S_{t},\nu_{t})$ where $O^{i}$ is solution on $[0,T^{i})\times \mathbb{R}^2_{+}$ of the following partial differential equation (PDE):
\begin{eqnarray}
0 &=& \partial_{t}O^{i}(t,S,\nu)+a_{\mathbb Q}(t,\nu)\partial_{\nu}O^{i}(t,S,\nu) \nonumber\\
&&+\frac{1}{2}\nu S^{2}\partial^{2}_{SS}O^{i}(t,S,\nu)+\rho\xi\nu  S \partial^{2}_{\nu S}O^{i}(t,S,\nu)+\frac{1}{2}\xi^{2}\nu\partial^{2}_{\nu\nu}O^{i}(t,S,\nu). \label{Heston PDE}
\end{eqnarray}

\begin{rem}
\label{remT}
  Options prices are also characterized by a terminal condition corresponding to the payoff. However, we will only consider short-term optimization problems for which the time horizon is before the maturity of all the options under consideration. Therefore, we shall never use the final condition associated with Eq. \eqref{Heston PDE}.\\
\end{rem}

\subsection{The optimization problem of the market maker}

We consider an option market maker in charge of providing bid and ask quotes for the $N$ above options over the period $[0,T]$ where $T< \text{min}_{i \in \{1, \ldots, N\}}T^{i}$ (see Remark \ref{remT}). For all $i \in \left\{1, \dots, N \right\}$, we denote by $\mathcal{O}^{i}_{t}-\delta_{t}^{i,b}(z)$ and $\mathcal{O}^{i}_{t}+\delta_{t}^{i,a}(z)$ the bid and ask prices (per contract) proposed by the market maker for a transaction corresponding to $z$ contracts of the $i$-th option, where $(\delta^{i}_{t}(.))_{t\in [0,T]}:=\big(\delta_{t}^{i,b}(.),\delta_{t}^{i,a}(.)\big)_{t\in [0,T]}$ is $\mathbb{F}$-predictable and bounded from below by a given constant $\delta_{\infty}$.\footnote{In applications, we always choose $\delta_{\infty}$ negative enough so that this lower bound is never binding.} Hereafter, we denote by $\mathcal{A}$ the set of $\mathbb{F}$-predictable $\mathbb R^{2N}$-valued maps that are bounded from below by $\delta_{\infty}$, and, by abuse of language, we call these maps admissible control processes. The dynamics of the inventory process $(q_{t})_{t\in [0,T]}:=(q_{t}^{1},\dots,q_{t}^{N})'_{t\in [0,T]}$ of the market maker is given by
\begin{align*}
dq_{t}^{i}:= \int_{\mathbb{R}_+^*} z \big(N^{i,b}(dt,dz)-N^{i,a}(dt,dz)\big), \forall i \in \{1, \ldots, N\},
\end{align*}
where, $\forall i \in \{1, \ldots, N\}$, $N^{i,b}(dt,dz)$ and $N^{i,a}(dt,dz)$ are two right-continuous $\mathbb{R}_+^*$-marked point processes, with almost surely no simultaneous jumps,\footnote{See Appendix \ref{A3} for more details on the construction of those processes.} modelling the transactions of the $i$-th option on the bid and ask side, whose respective intensity processes $(\lambda_{t}^{i,b}(dz))_{t\in \mathbb{R}_{+}}$ and $(\lambda_{t}^{i,a}(dz))_{t\in \mathbb{R}_{+}}$ are given by
\begin{equation*}
\lambda^{i,b}_{t}(dz):=\Lambda^{i,b}(\delta_{t}^{i,b}(z))\mathds{1}_{\{q_{t-}+z e^{i} \in \mathcal{Q}\}} \mu^{i,b}(dz)  \qquad \lambda^{i,a}_{t}(dz):=\Lambda^{i,a}(\delta_{t}^{i,a}(z))\mathds{1}_{\{q_{t-}-z e^{i} \in \mathcal{Q}\}} \mu^{i,a}(dz)
\end{equation*}
with $(e^{i})_{i \in \{1, \ldots, N\}}$ the canonical basis of $\mathbb{R}^{N}$, $\mathcal{Q}$ the set of authorized inventories\footnote{The frontier of this set defines the risk limits of the market maker.} for the market maker, and $(\mu^{i,b}, \mu^{i,a})$ a couple of probability measures on $\mathbb{R}_+^*$ modelling the distributions of transaction sizes. For $i \in \left\{1, \dots, N \right\}$, $\Lambda^{i,b}$ and $\Lambda^{i,a}$ are positive functions satisfying the following classical hypotheses (see~\cite{gueant2016financial, gueant2017optimal} for similar assumptions):
\begin{itemize}
\item $\Lambda^{i,b}$ and $\Lambda^{i,a}$ are twice continuously differentiable.
\item $\Lambda^{i,b}$ and $\Lambda^{i,a}$ are strictly decreasing, with $\Lambda^{i,b'}<0$ and $\Lambda^{i,a'}<0$.
\item $\underset{\delta\rightarrow +\infty}{\text{lim}}\Lambda^{i,b}(\delta)=\underset{\delta\rightarrow +\infty}{\text{lim}}\Lambda^{i,a}(\delta)=0$.
\item $\underset{\delta\in \mathbb{R}}{\text{sup}} \frac{\Lambda^{i,b}(\delta)\Lambda^{i,b''}(\delta)}{\big(\Lambda^{i,b'}(\delta)\big)^{2}}<2$ and $\underset{\delta\in \mathbb{R}}{\text{sup}} \frac{\Lambda^{i,a}(\delta)\Lambda^{i,a''}(\delta)}{\big(\Lambda^{i,a'}(\delta)\big)^{2}}<2$.\\
\end{itemize}
The above conditions are sufficiently general to allow for the vast majority of relevant forms of intensities: the exponential intensities initially introduced in~\cite{avellaneda2008high} and used in most of the literature, logistic intensities as in~\cite{bergault2019optimal}, or many SU Johnson intensities as in~\cite{gueant2019deep}.\\

In addition to quoting prices for the $N$ options, the market maker can buy and sell the underlying asset. We assume that the market for that asset is liquid enough to ensure a perfect $\Delta$-hedging.\\

\begin{rem}
In practice, for a portfolio that is not vega-hedged, it is usually suboptimal to perfectly $\Delta$-hedge the portfolio because of the correlation between the spot process and the instantaneous variance process. Nevertheless, we assume here for the sake of simplicity that $\Delta$-hedging is carried out in continuous time. A study of the optimal position in the underlying asset and its consequence on our problem is carried out in Appendix \ref{A1}.\\
\end{rem}

In what follows, we denote by $(\Delta_{t})_{t\in [0,T]}$ the $\Delta$ of the portfolio:
\begin{equation*}
\Delta_{t}:=\underset{i=1}{\overset{N}{\sum}}\partial_{S}O^{i}(t,S_{t},\nu_{t})q_{t}^{i}    \text{ for all } t\in [0,T].
\end{equation*}
The resulting dynamics for the cash process $(X_{t})_{t\in [0,T]}$ of the market maker is:
\begin{align*}
dX_{t}:=\underset{i=1}{\overset{N}{\sum}}\bigg(\int_{ \mathbb{R}_+^*}z\Big(\delta_{t}^{i,b}(z)N^{i,b}(dt,dz)+\delta_{t}^{i,a}(z)N^{i,a}(dt,dz)\Big)-\mathcal{O}_{t}^{i}dq_{t}^{i}\bigg)+S_{t}d\Delta_{t}+d\big\langle \Delta,S\big\rangle_t.
\end{align*}
We denote by $(V_{t})_{t\in [0,T]}$ the process for the Mark-to-Market (MtM) value of the market maker's portfolio (cash, shares, and options), i.e.,
\begin{align*}
V_{t}:=X_{t}-\Delta_{t}S_{t}+\underset{i=1}{\overset{N}{\sum}}q_{t}^{i}\mathcal{O}_{t}^{i}.
\end{align*}
The dynamics of that process is given by
\begin{eqnarray*}
 dV_{t} &=& dX_{t}-S_{t}d\Delta_{t} - \Delta_{t}dS_{t} - d\big\langle \Delta,S\big\rangle_t  +\underset{i=1}{\overset{N}{\sum}} \mathcal{O}_{t}^{i}dq_{t}^{i} +\underset{i=1}{\overset{N}{\sum}} q_{t}^{i}d\mathcal{O}_{t}^{i}\\
 \!\!\!&\!\!=\!\!&\!\!\!\underset{i=1}{\overset{N}{\sum}}\left(\int_{\mathbb{R}_+^*}z\Big(\delta_{t}^{i,b}(z)N^{i,b}(dt,dz)+\delta_{t}^{i,a}(z)N^{i,a}(dt,dz)\Big)+q_{t}^{i}d\mathcal{O}_{t}^{i}\right)-\Delta_{t}dS_{t} \\
\!\!\!&\!\!=\!\!&\!\!\! \underset{i=1}{\overset{N}{\sum}} \Bigg(\int_{\mathbb{R}_+^*} z\Big(\delta_{t}^{i,a}(z)N^{i,a}(dt,dz)+\delta_{t}^{i,b}(z)N^{i,b}(dt,dz)\Big)\!+\!q_{t}^{i}\partial_{\nu}O^{i}(t,S_{t},\nu_{t})\big( a_{\mathbb{P}}(t,\nu_t) - a_{\mathbb{Q}}(t,\nu_t) \big)dt\\
&&\qquad \qquad \quad \qquad \qquad \!+\!\sqrt{\nu_{t}}\xi q_{t}^{i}\partial_{\nu}O^{i}(t,S_{t},\nu_{t})dW_{t}^{\nu}\Bigg).
\end{eqnarray*}

For all $i \in \left\{1, \dots, N \right\}$, the vega of the $i$-th option is defined as
\begin{align*}
\mathcal{V}_{t}^{i}:=\partial_{\sqrt{\nu}}O^{i}(t,S_{t},\nu_{t})=2\sqrt{\nu_{t}}\partial_{\nu}O^{i}(t,S_{t},\nu_{t}) \text{ for all } t\in[0,T].
\end{align*}

Therefore, we can rewrite the dynamics of the portfolio as
\begin{align*}
& dV_{t}=\underset{i=1}{\overset{N}{\sum}}\left(\int_{\mathbb{R}_+^*}z\Big(\delta_{t}^{i,b}(z)N_{t}^{i,b}(dt,dz)+\delta_{t}^{i,a}(z)N_{t}^{i,a}(dt,dz)\Big)+q_{t}^{i} \mathcal{V}_{t}^{i}\frac{a_{\mathbb{P}}(t,\nu_t) - a_{\mathbb{Q}}(t,\nu_t)}{2\sqrt{\nu_{t}}}dt+\frac{\xi}{2} q_{t}^{i}\mathcal{V}_{t}^{i}dW_{t}^{\nu}\right).
\end{align*}

Following the academic literature on market making, we can consider two objective functions. As in the initial Avellaneda and Stoikov setting~\cite{avellaneda2008high} (see also~\cite{gueant2016financial,gueant2017optimal,gueant2013dealing}), we can consider the following expected utility objective function:
\begin{equation*}
\sup_{\delta\in \mathcal{A}}\mathbb{E}\bigg[-\exp\big(-\gamma V_{T}\big)\bigg],
\end{equation*}
where $\gamma>0$ is the risk--aversion parameter of the market maker.
Instead, as in~\cite{cartea2017algorithmic,cartea2015algorithmic,cartea2014buy}, but also in~\cite{gueant2017optimal}, we can consider a risk--adjusted expectation for the objective function,~i.e.
\begin{equation*}
\sup_{\delta\in \mathcal{A}}\mathbb{E}\left[V_{T}-\frac{\gamma}{2}\int_{0}^{T}\left(\underset{i=1}{\overset{N}{\sum}}\frac{\xi}{2} q_{t}^{i}\mathcal{V}_{t}^{i}\right)^{2}dt\right].
\end{equation*}
The second objective function in our case writes
\begin{align}
\sup_{\delta\in \mathcal{A}}\mathbb{E}&\left[\int_{0}^{T}\left(\underset{i=1}{\overset{N}{\sum}}\left(\left(\underset{j=a,b}{\sum}\int_{\mathbb{R}_+^*}z\delta_{t}^{i,j}(z)\Lambda^{i,j}(\delta_{t}^{i,j}(z))\mathds{1}_{\{q_{t-}-\psi(j)ze^{i} \in \mathcal{Q}\}} \mu^{i,j}(dz)\right)\right.\right.\right.\nonumber\\
&\quad +\left.\left.\left.\vphantom{\int_{0}^{T}\left(\underset{i=1}{\overset{N}{\sum}}\left(\left(\underset{j=a,b}{\sum}\int_{\mathbb{R}_+^*}z\delta_{t}^{i,j}(z)\Lambda^{i,j}(\delta_{t}^{i,j}(z))\mathds{1}_{\{q_{t-}-\psi(j)ze^{i} \in \mathcal{Q}\}} \mu^{i,j}(dz)\right)\right.\right.}q_{t}^{i}\mathcal{V}_{t}^{i} \frac{a_{\mathbb{P}}(t,\nu_t) - a_{\mathbb{Q}}(t,\nu_t)}{2\sqrt{\nu_{t}}}\right)\right)dt-\frac{\gamma\xi^{2}}{8}\int_{0}^{T}\left(\underset{i=1}{\overset{N}{\sum}}q_{t}^{i}\mathcal{V}_{t}^{i}\right)^{2}dt\right],\label{Market Maker Problem Risk Adjusted Model}
\end{align}
where
\begin{align*}
\psi(j):=\left\{
    \begin{array}{ll}
        +1 \text{ if } j=a  \\
        -1 \text{ if } j=b.
    \end{array}
\right.
\end{align*}

These two objective functions are close to one other in practice. Guéant showed in \cite{gueant2017optimal} that they give similar optimal quotes in practical examples. Furthermore, in many cases, the expected utility framework with exponential utility function can be reduced to the maximization of the expected PnL minus a quadratic penalty of the above form, up to a change in the intensity functions (see \cite{manziuk}).\\

In what follows, we consider the second framework. Therefore, we define the value function $$u : \big(t,S, \nu,q\big) \in [0,T] \times \mathbb{R}_{+}^2 \times \mathcal{Q} \mapsto u\big(t,S,\nu,q\big)$$ associated with \eqref{Market Maker Problem Risk Adjusted Model} as

\begin{eqnarray*}
u\big(t,S, \nu,q\big)  &=& \sup_{(\delta_s)_{s \in [t,T]}\in \mathcal{A}_t} \mathbb{E}_{(t,S,\nu,q)}\left[\int_{t}^{T}\underset{i=1}{\overset{N}{\sum}}\left(\left(\underset{j=a,b}{\sum}\int_{\mathbb{R}_+^*}z\delta_{s}^{i,j}(z)\Lambda^{i,j}(\delta_{s}^{i,j}(z))\mathds{1}_{\{q_{s-}-\psi(j)ze^{i} \in \mathcal{Q}\}} \mu^{i,j}(dz)\right)\right.\right.\\
&&\left.\left.{}\vphantom{\sup_{\delta\in \mathcal{A}_t} \mathbb{E}_{(t,S,\nu,q)}\left[\int_{t}^{T}\underset{i=1}{\overset{N}{\sum}}\left(\left(\underset{j=a,b}{\sum}z^{i}\delta_{s}^{i,j}\Lambda^{i,j}(\delta_{s}^{i,j})\right)\right.\right. }+q_{s}^{i}\mathcal{V}_{s}^{i} \frac{a_{\mathbb{P}}(s,\nu_s) - a_{\mathbb{Q}}(s,\nu_s)}{2\sqrt{\nu_{s}}}\right) ds-\frac{\gamma\xi^{2}}{8}\int_{t}^{T}\left(\underset{i=1}{\overset{N}{\sum}}q_{s}^{i}\mathcal{V}_{s}^{i}\right)^{2}ds\right],
\end{eqnarray*}
where $\mathcal{A}_t$ is the set of admissible controls ($\mathbb{F}$-predictable $\mathbb R^{2N}$-valued maps that are bounded from below by~$\delta_{\infty}$) defined on $[t,T]$.\\

\subsection{Assumptions and approximations}

The above stochastic optimal control problem can be addressed from a theoretical point of view using an approach similar to that of \cite{gueant2017optimal}. However, when it comes to approximating the optimal quotes a market maker should set for the $N$ options, classic numerical methods are of no help because the value function $u$ has $N+2$ variables (in addition to the time variable). In order to beat the curse of dimensionality we propose a method based on the following assumptions/approximations:

\begin{assumption}
We approximate the vega of each option over $[0,T]$ by its value at time $t=0$, namely
\begin{align*}
\mathcal{V}_{t}^{i} = \mathcal{V}_{0}^{i} =:\mathcal{V}^{i}\in \mathbb{R}, \text{ for all } i \in \{1, \ldots, N\}.
\end{align*}
\end{assumption}

\begin{assumption}
We assume that the set of authorized inventories is associated with vega risk limits, i.e.
\begin{equation*}
\mathcal{Q}=\bigg\{q\in \mathbb{R}^{N} \left| \underset{i=1}{\overset{N}{\sum}}q^{i}\mathcal{V}^{i}\in \left[-\overline{\mathcal{V}},\overline{\mathcal{V}}\right]  \right.\bigg\},
\end{equation*}
where $\overline{\mathcal{V}}\in \mathbb{R}_+^*$ is the vega risk limit of the market maker.\\
\end{assumption}

The first assumption is acceptable if $T$ is not too large. This raises in fact the deep question of the reasonable value of $T$, as there is no natural choice for the horizon of the optimization problem. In practice, $T$ has to be sufficient large to allow for several transactions in many options and small enough for the constant-vega approximation to be relevant (and smaller than the maturities of the options). It is also noteworthy, although it is time-inconsistent, that one can use the output of the model (with the constant-vega approximation) over a short period of time and then run the model again with updated vegas. This is a classical practice in applied optimal control when the parameters are estimated online.\footnote{In Appendix \ref{A2} we propose a method to relax the constant-vega assumption. This method is based on a Taylor expansion around the constant-vega case. The curse of dimensionality is tamed by the reduction of the problem to a Monte-Carlo simulation.}\\

The second assumption states that risk limits are related to the only source of risk (as the portfolio is $\Delta$-hedged). This is a natural assumption. The only drawback is that no risk limit can be set to individual options.\\

\section{An approximate solution to the problem}

\subsection{Change of variables: beating the curse of dimensionality}

Under the above assumptions, the $N+2$ state variables can be replaced by only two: the instantaneous variance and the vega of the portfolio. This portfolio vega, defined by $\mathcal{V}_t^\pi := \sum_{i=1}^N q_t^i \mathcal{V}^i$, has the following dynamics:
$$d\mathcal{V}^\pi_t = \sum_{i=1}^N \int_{\mathbb{R}_+^*}z\mathcal{V}^i\big(N^{i,b}(dt,dz)-N^{i,a}(dt,dz)\big).$$

It is clear then that the value function $u$ verifies
$$ \forall (t,S,\nu,q\big) \in  [0,T] \times \mathbb{R}_{+}^2 \times \mathcal{Q}, u(t,S,\nu,q\big) = v\left(t,\nu,\sum_{i=1}^N q^i \mathcal{V}^i\right),$$
where
\begin{equation}
\begin{split}
v\left(t,\nu,\mathcal{V}^\pi\right)  = \sup_{(\delta_s)_{s \in [t,T]}\in \mathcal{A}_t} \mathbb{E}_{(t,\nu,\mathcal{V}^\pi)}\left[ \vphantom{\underset{i=1}{\overset{N}{\sum}}\underset{j=a,b}{\sum}\int_{\mathbb{R}_+^*}z\delta_{s}^{i,j}(z)\Lambda^{i,j}(\delta_{s}^{i,j}(z))\mathds{1}_{\left\{|\mathcal{V}^{\pi}_s-\psi(j)z\mathcal{V}^{i}|\leq \overline{\mathcal V}\right\}}} \int_{t}^{T} \right.&\left(\left(\underset{i=1}{\overset{N}{\sum}}\underset{j=a,b}{\sum} \int_{\mathbb{R}_+^*}z\delta_{s}^{i,j}(z)\Lambda^{i,j}(\delta_{s}^{i,j}(z))\mathds{1}_{\left\{|\mathcal{V}^{\pi}_s-\psi(j)z\mathcal{V}^{i}|\leq \overline{\mathcal V}\right\}} \mu^{i,j}(dz)\right) \right.\\
& \left. \left.+ \mathcal{V}^{\pi}_s \frac{a_{\mathbb{P}}(s,\nu_s) - a_{\mathbb{Q}}(s,\nu_s)}{2\sqrt{\nu_{s}}} -\frac{\gamma\xi^{2}}{8} {\mathcal{V}^\pi_s}^2 \vphantom{\underset{i=1}{\overset{N}{\sum}}\underset{j=a,b}{\sum}z^{i}\delta_{s}^{i,j}\Lambda^{i,j}(\delta_{s}^{i,j})\mathds{1}_{\left\{|\mathcal{V}^{\pi}_s-\psi(j)z^{i}\mathcal{V}^{i}|\leq \overline{\mathcal V}\right\}}} \right) ds\right].
\end{split}
\label{v_def}
\end{equation}

Because $(\nu_t, \mathcal{V}^\pi_t)_{t \in [0,T]}$ is a Markov process, the problem boils down, under the two above assumptions, to a low-dimensional optimal control problem where the two state variables are driven by $2N$ controlled point processes and a standard Brownian motion.\\

\subsection{Hamilton-Jacobi-Bellman equation and optimal controls}

Following \cite{sulem}, the Hamilton-Jacobi-Bellman equation associated with \eqref{v_def} is given by
\begin{equation}\label{Reduced HJB equation}
\begin{split}
0=\ &\partial_{t}v(t,\nu,\mathcal{V}^\pi) + a_{\mathbb{P}}(t,\nu) \partial_{\nu} v(t,\nu,\mathcal{V}^{\pi}) + \frac 12 \nu \xi^2 \partial^2_{\nu \nu} v(t,\nu,\mathcal{V}^{\pi}) + \mathcal{V}^{\pi}\frac{a_{\mathbb{P}}(t,\nu) - a_{\mathbb{Q}}(t,\nu)}{2\sqrt{\nu}} -\frac{\gamma\xi^{2}}{8}{\mathcal{V}^{\pi}}^{2}\\
&+\underset{i=1}{\overset{N}{\sum}}\underset{j=a,b}{\sum}\int_{\mathbb{R}_+^*}z\mathds{1}_{\left\{|\mathcal{V}^{\pi}-\psi(j)z\mathcal{V}^{i}|\leq \overline{\mathcal V}\right\}}H^{i,j}\bigg(\frac{v\big(t,\nu,\mathcal{V}^{\pi}\big)-v\big(t,\nu,\mathcal{V}^{\pi} -\psi(j)z\mathcal{V}^{i}\big)}{z}\bigg) \mu^{i,j}(dz),
\end{split}
\end{equation}
with final condition $v(T,\nu,\mathcal{V}^\pi) = 0$, where
\begin{align*}
H^{i,j}(p):=\sup_{\delta^{i,j} \ge \delta_\infty}\Lambda^{i,j}(\delta^{i,j})(\delta^{i,j}-p), \text{ }i \in \{1, \ldots, N\}, \text{ } j=a,b.
\end{align*}

We end up therefore with a low-dimensional functional equation of the Hamilton-Jacobi-Bellman type.\\

Once the value function is known, the optimal controls, which are the optimal mid-to-bid and ask-to-mid associated with the $N$ options, are given by the following formula (see \cite{bergault2019optimal, gueant2017optimal}):
$$\delta^{i,j*}_t(z) = \max\left(\delta_\infty, \left(\Lambda^{i,j}\right)^{-1} \left( -H^{i,j'}\left(\frac{v\big(t,\nu_t,\mathcal{V}_{t-}^{\pi}\big)-v\big(t,\nu_t,\mathcal{V}_{t-}^{\pi} -\psi(j)z\mathcal{V}^{i}\big)}{z}\right) \right)\right), \text{ }i \in \{1, \ldots, N\}, \text{ } j=a,b.$$

\begin{rem}
In the case where $a_{\mathbb{P}} = a_{\mathbb{Q}}$, it is evident that $v$ does not depend on $\nu$. In that case indeed, we have $v(t,\nu,\mathcal{V}^\pi) = w(t,\mathcal{V}^\pi)$ where $w$ is solution of the simpler Hamilton-Jacobi-Bellman equation
\begin{equation*}
0= \partial_{t}w(t,\mathcal{V}^\pi) -\frac{\gamma\xi^{2}}{8}{\mathcal{V}^{\pi}}^{2} +\underset{i=1}{\overset{N}{\sum}}\underset{j=a,b}{\sum}\int_{\mathbb{R}_+^*}z\mathds{1}_{\left\{|\mathcal{V}^{\pi}-\psi(j)z\mathcal{V}^{i}|\leq \overline{\mathcal V}\right\}}H^{i,j}\bigg(\frac{w\big(t,\mathcal{V}^{\pi}\big)-w\big(t,\mathcal{V}^{\pi} -\psi(j)z\mathcal{V}^{i}\big)}{z}\bigg) \mu^{i,j}(dz),
\end{equation*}
with final condition $w(T,\mathcal{V}^\pi) = 0$.
\end{rem}

\section{Numerical results}

\subsection{Model parameters}

In this section we consider a book of options and derive the optimal quotes using the above approach.\\

For this purpose, we consider an underlying stock with the following characteristics:
\begin{itemize}
    \item[-] Stock price at time $t=0$: $S_0=10\ \text{\euro}.$
    \item[-] Instantaneous variance at time $t=0$: $\nu_0=0.0225\text{ year}^{-1}$.
    \item[-] Heston model with $a_{\mathbb{P}}(t,\nu) = \kappa_{\mathbb{P}} (\theta_\mathbb{P} - \nu)$ where $\kappa_{\mathbb{P}} = 2 \text{ year}^{-1}$ and $\theta_{\mathbb{P}} = 0.04 \text{ year}^{-1}$, and $a_{\mathbb{Q}}(t,\nu) = \kappa_{\mathbb{Q}} (\theta_\mathbb{Q} - \nu)$ where $\kappa_{\mathbb{Q}} = 3 \text{ year}^{-1}$ and $\theta_{\mathbb{Q}} = 0.0225 \text{ year}^{-1}$.
    \item[-] Volatility of volatility parameter: $\xi=0.2\text{ year}^{-1}$.
    \item[-] Spot-variance correlation: $\rho=-0.5$.
\end{itemize}

We consider the case of a market maker dealing with 20 European call options written on that stock where the strike$\times$maturity couples are the elements $(K^i,T^i)_{i=1, \ldots, 20}$ of the set $\mathcal{K}\times\mathcal{T}$, where
\begin{align*}
    \mathcal{K}=\{8\ \text{\euro},9\ \text{\euro},10\ \text{\euro},11\ \text{\euro},12\ \text{\euro} \}\quad \text{and} \quad \mathcal{T}=\{1\text{ year},1.5\text{ years},2\text{ years},3\text{ years} \}.
\end{align*}

\begin{figure}[!h]
\begin{center}
\includegraphics[width=\textwidth]{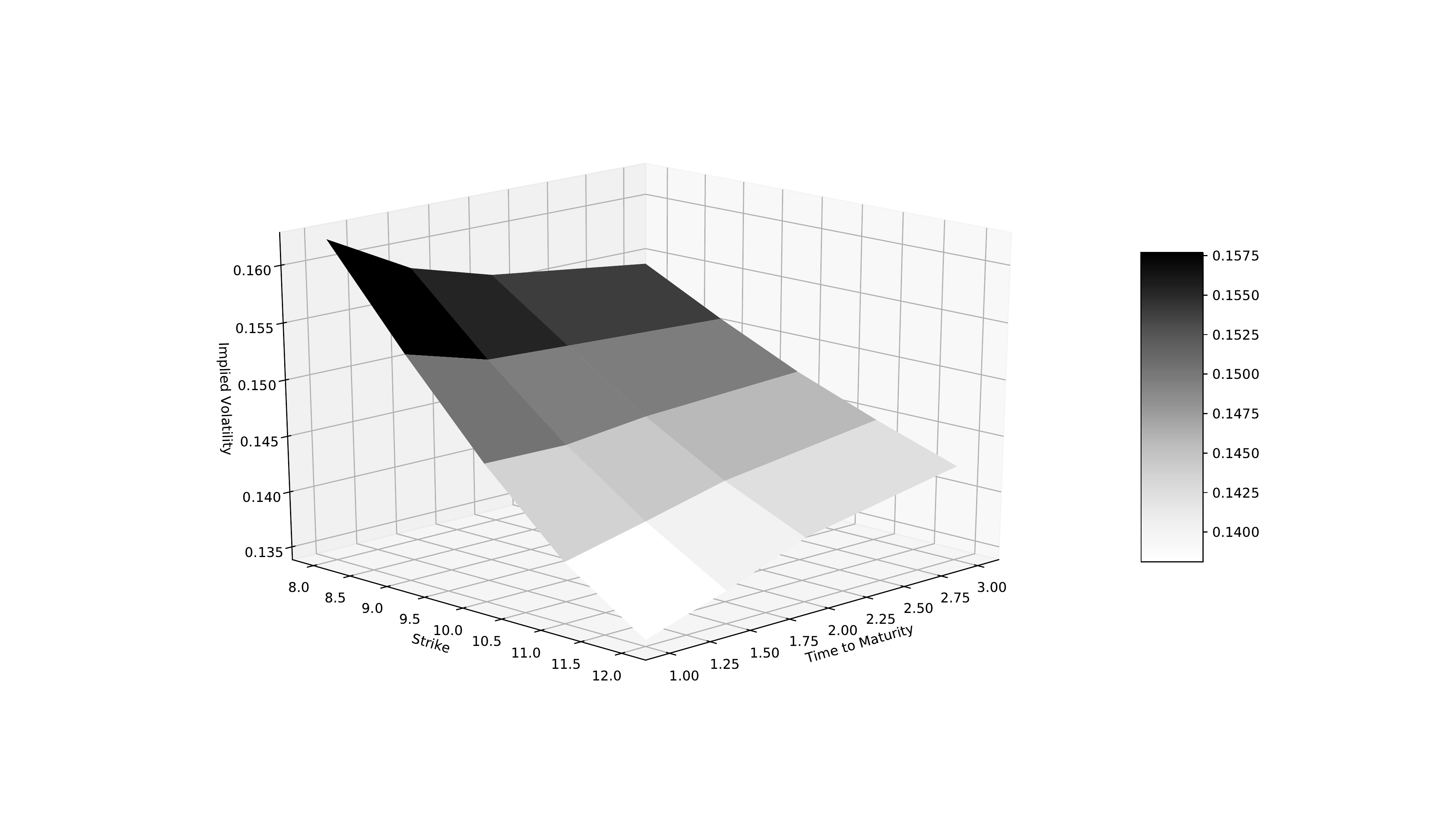}
\caption{Implied volatility surface associated with the above parameters}
\label{Vol surface Heston}
\end{center}
\end{figure}

The associated implied volatility surface is plotted in Figure \ref{Vol surface Heston}.\footnote{This plot has been computed using $10^5$ Monte-Carlo simulations for each option.}\\

As far as the liquidity parameters of these options are concerned, we consider the following intensity functions:
\begin{align*}\Lambda^{i,j}(\delta)=\frac{\lambda^i}{1+e^{\alpha +\frac{\beta}{\mathcal{V}^i} \delta}}, \text{ }i \in \{1, \ldots, N\}, \text{ } j=a,b,
\end{align*}
where $\lambda^i= \frac{252 \times 30}{1 + 0.7 \times |S_0 - K^i|}\text{ year}^{-1}$, $\alpha=0.7$, and $\beta=150 \text{ year}^{\frac 12}$. The choice of $\lambda^i$  corresponds to $30$ requests per day for at-the-money options, and decreases to 12.5 for the most in- and out-the-money options. The choice of $\alpha$ corresponds to a probability of $\frac{1}{1+e^{0.7}}\approx 33\%$ to trade when the answered quote is the mid-price. The choice of $\beta$ corresponds to a probability of $\frac{1}{1+e^{-0.8}}\approx 69\%$ to trade when the answered quote corresponds to an implied volatility $1\%$ better for the client and a probability of $\frac{1}{1+e^{2.2}}\approx 10\%$ to trade when the answered quote corresponds to an implied volatility $1\%$ worse for the client.\\

Moreover, we assume transactions of constant size with $z^i=\frac{5\cdot 10^5}{\mathcal{O}_0^i}\text{ contracts}$ for option $i$. This corresponds approximately to $500000\text{ \text{\euro}}$ per transaction.\footnote{This is only an approximation as trade sizes are in number of options and option prices move.} In other words, the measures $\mu^{i,b}$ and $\mu^{i,a}$ are Dirac masses at $z^i$.\\

Regarding the vega risk limits, we consider $\overline{\mathcal V}=10^7\text{ \text{\euro}}\cdot \text{year}^{\frac{1}{2}}$.\\

As far as the time horizon is concerned, we consider $T=0.0012$ year (i.e. $0.3$ day). This short time horizon surprisingly ensures convergence towards stationary quotes at time $t=0$ (see Figure \ref{quotes_conv} below).\\

Eventually, we consider a risk aversion parameter $\gamma=1\cdot 10^{-3}\text{ \text{\euro}}^{-1}$.\\

\subsection{Optimal quotes}

\begin{figure}[!h]
\begin{center}
\includegraphics[width=1.03\textwidth]{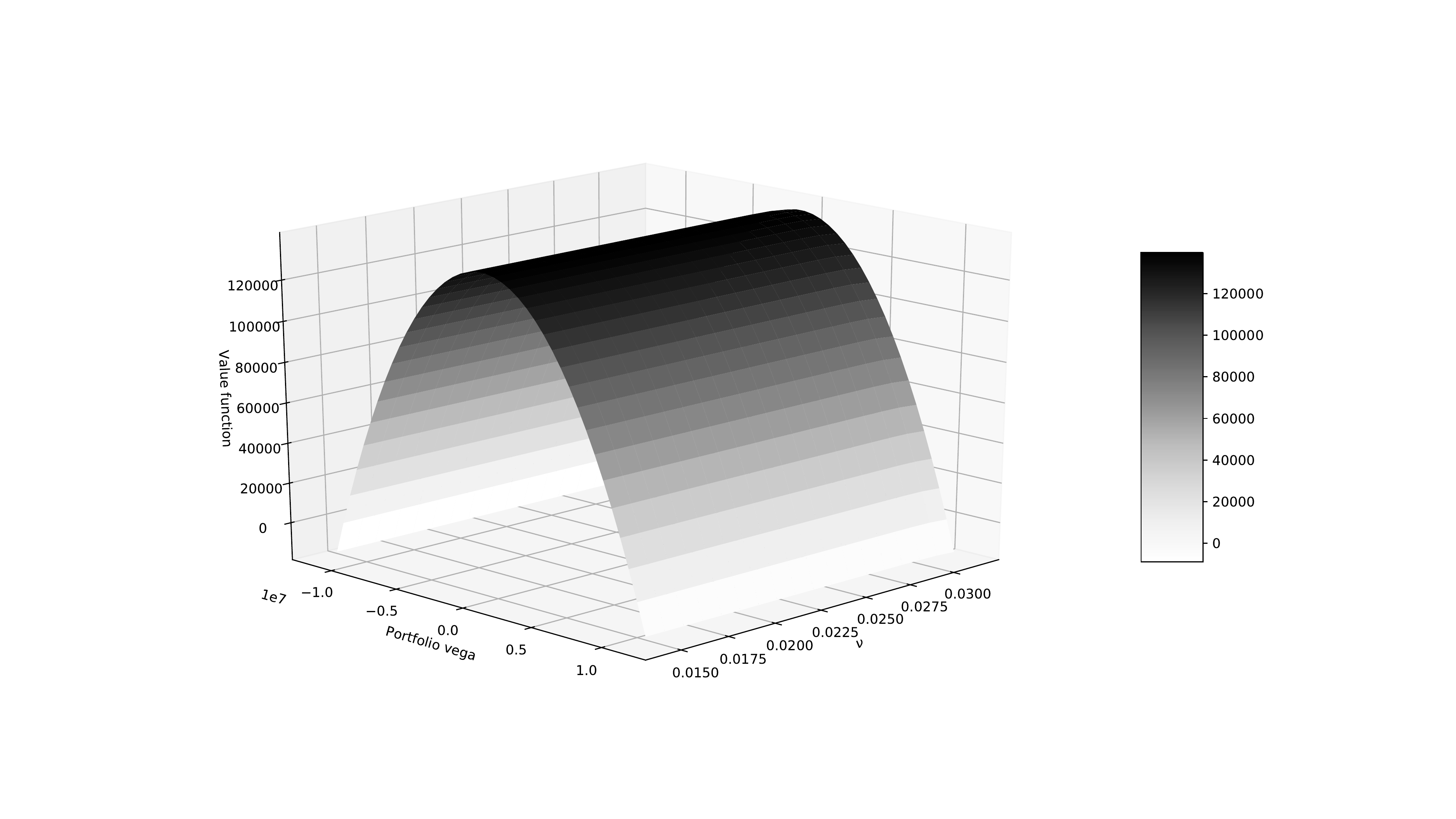}
\caption{Value function as a function of instantaneous variance and portfolio vega.}
\label{Value_function_vega}
\end{center}
\end{figure}

Using a monotone explicit Euler scheme with linear interpolation on a grid of size $180\times 30 \times 40$, we approximate the value function solution to \eqref{Reduced HJB equation} (with Neumann conditions at the boundaries in $\nu$) on the domain $[0,T]\times[0.0144,0.0324]\times\left[-\overline{\mathcal{V}}, \overline{\mathcal{V}}\right]$. This value function is plotted in Figure \ref{Value_function_vega}. We see that the dependence in $\nu$ is low. This is not surprising given that, at the time scale of the market making problem, the discrepancy between $a_{\mathbb{P}}$ and $a_{\mathbb{Q}}$ could be ignored.\\

\begin{figure}[!h]
\begin{center}
\includegraphics[width=\textwidth]{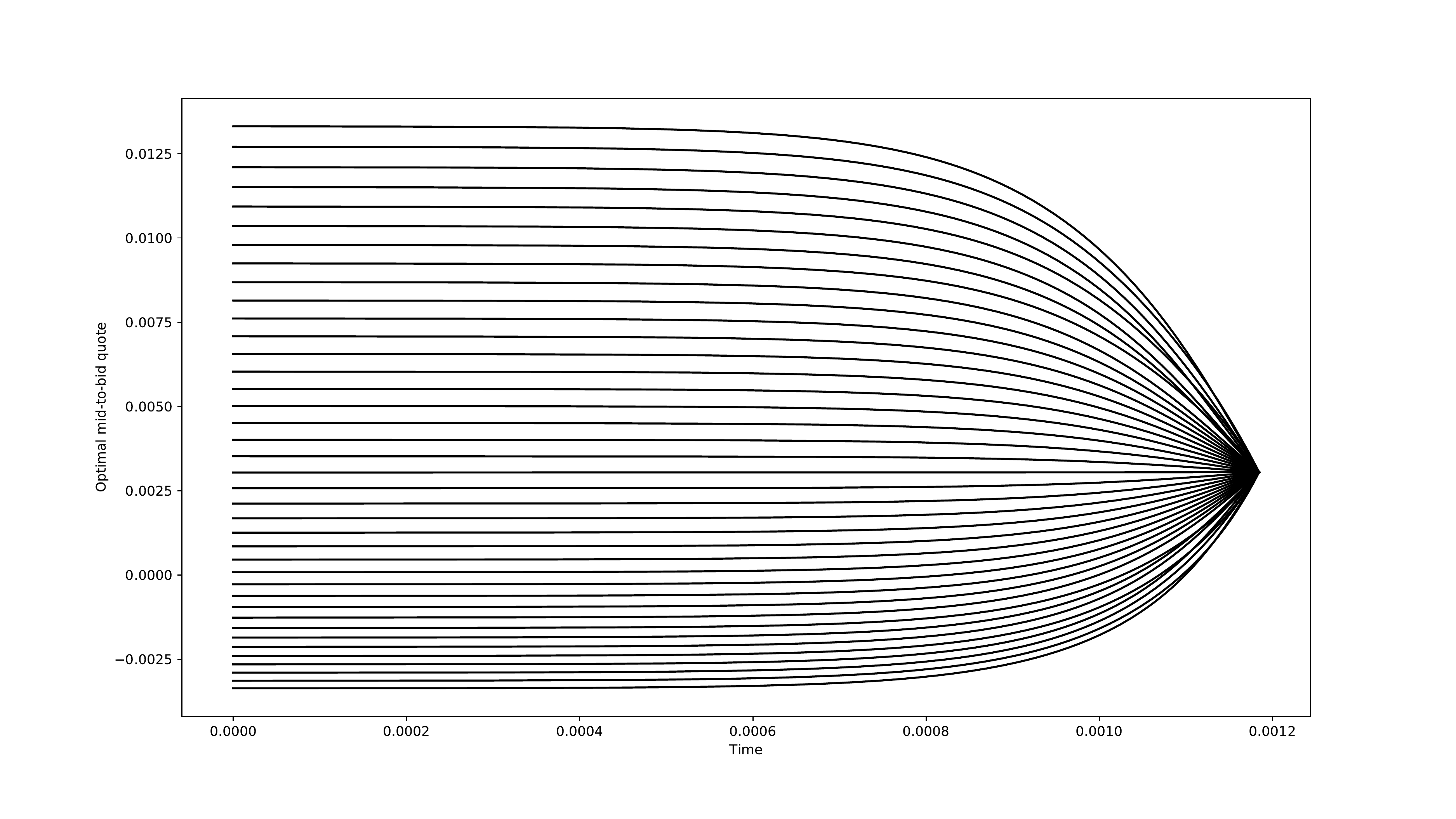}
\caption{Optimal mid-to-bid quotes as a function of time for option 1: $(K^1,T^1) = (8,1)$ -- $\nu = 0.0225$.}
\label{quotes_conv}
\end{center}
\end{figure}

From that value function, we deduce the optimal bid and ask quotes of the market maker (in fact mid-to-bid and ask-to-mid) for each option as a function of the portfolio vega. As mentioned above, we chose $T=0.0012$ year (i.e. $0.3$ day) -- a choice that ensures convergence of the optimal quotes to their stationary values (see Figure \ref{quotes_conv}).\\

\begin{figure}[!h]
\begin{center}
\includegraphics[width=0.92\textwidth]{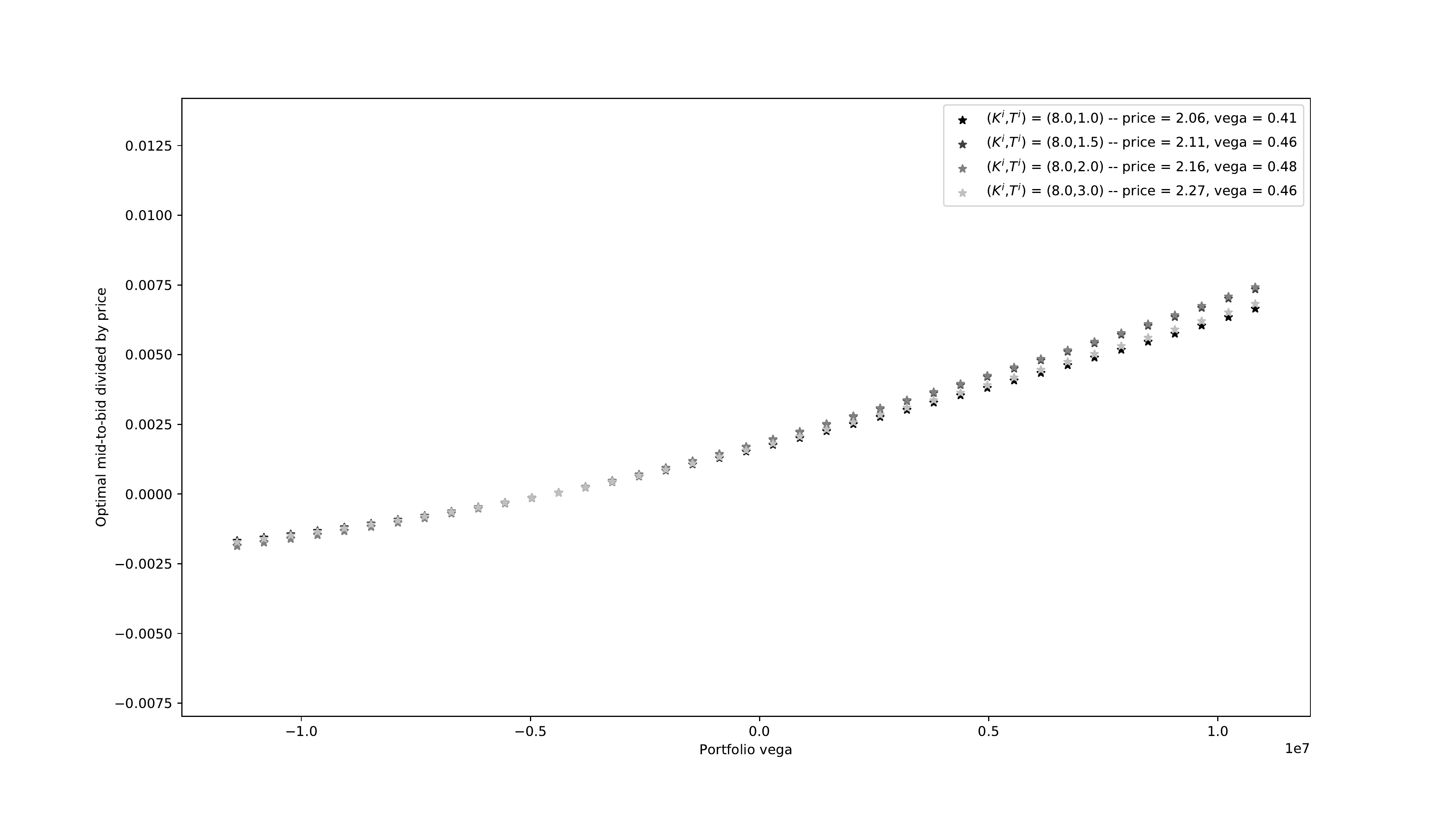}
\caption{Optimal mid-to-bid quotes divided by option price as a function of the portfolio vega for K=8.}
\label{optimal_quotes_8}
\end{center}
\end{figure}

\begin{figure}[!h]
\begin{center}
\includegraphics[width=0.92\textwidth]{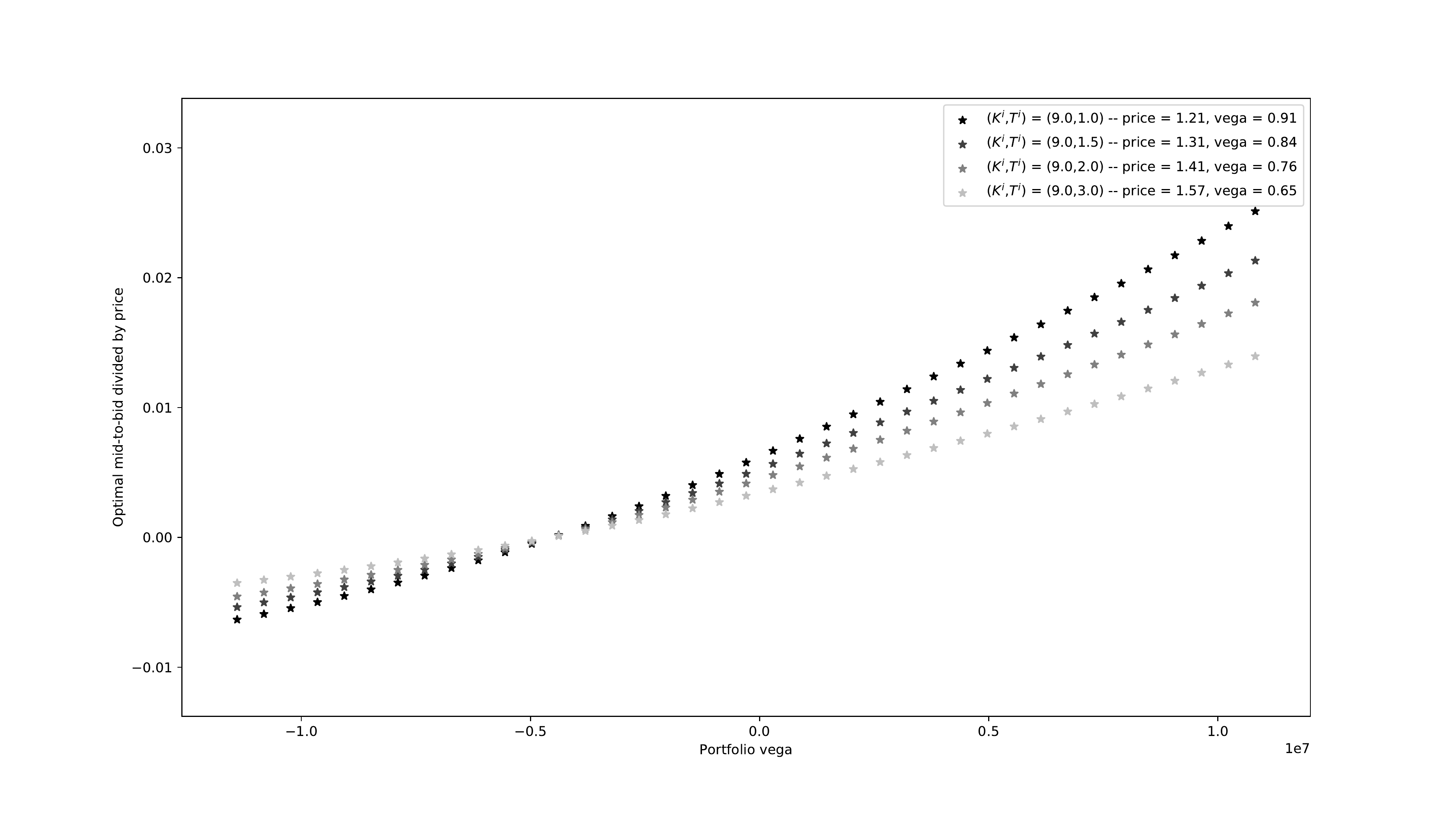}
\caption{Optimal mid-to-bid quotes divided by option price as a function of the portfolio vega for K=9.}
\label{optimal_quotes_9}
\end{center}
\end{figure}

\begin{figure}[!h]
\begin{center}
\includegraphics[width=0.92\textwidth]{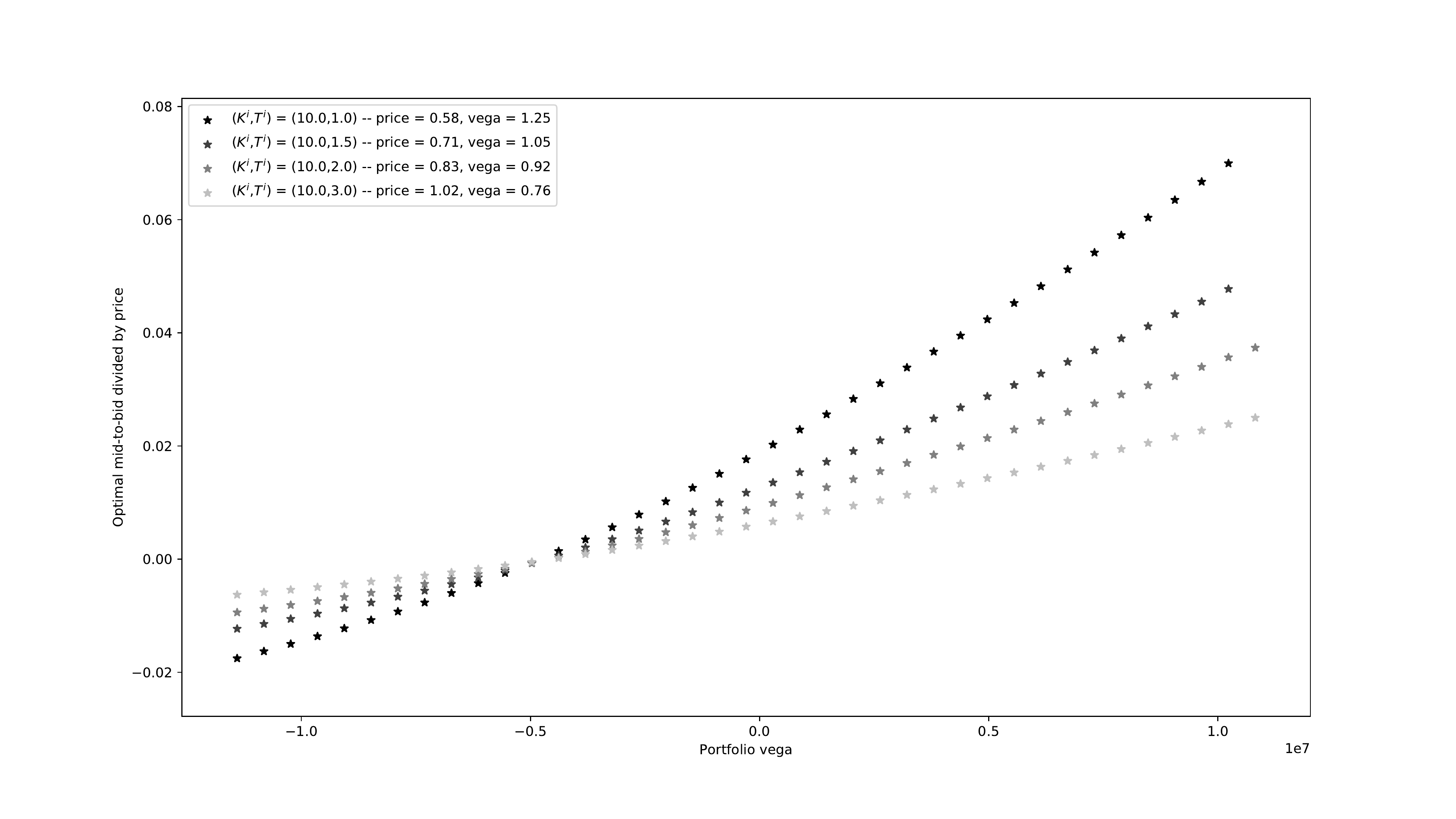}
\caption{Optimal mid-to-bid quotes divided by option price as a function of the portfolio vega for K=10.}
\label{optimal_quotes_10}
\end{center}
\end{figure}

\begin{figure}[!h]
\begin{center}
\includegraphics[width=0.92\textwidth]{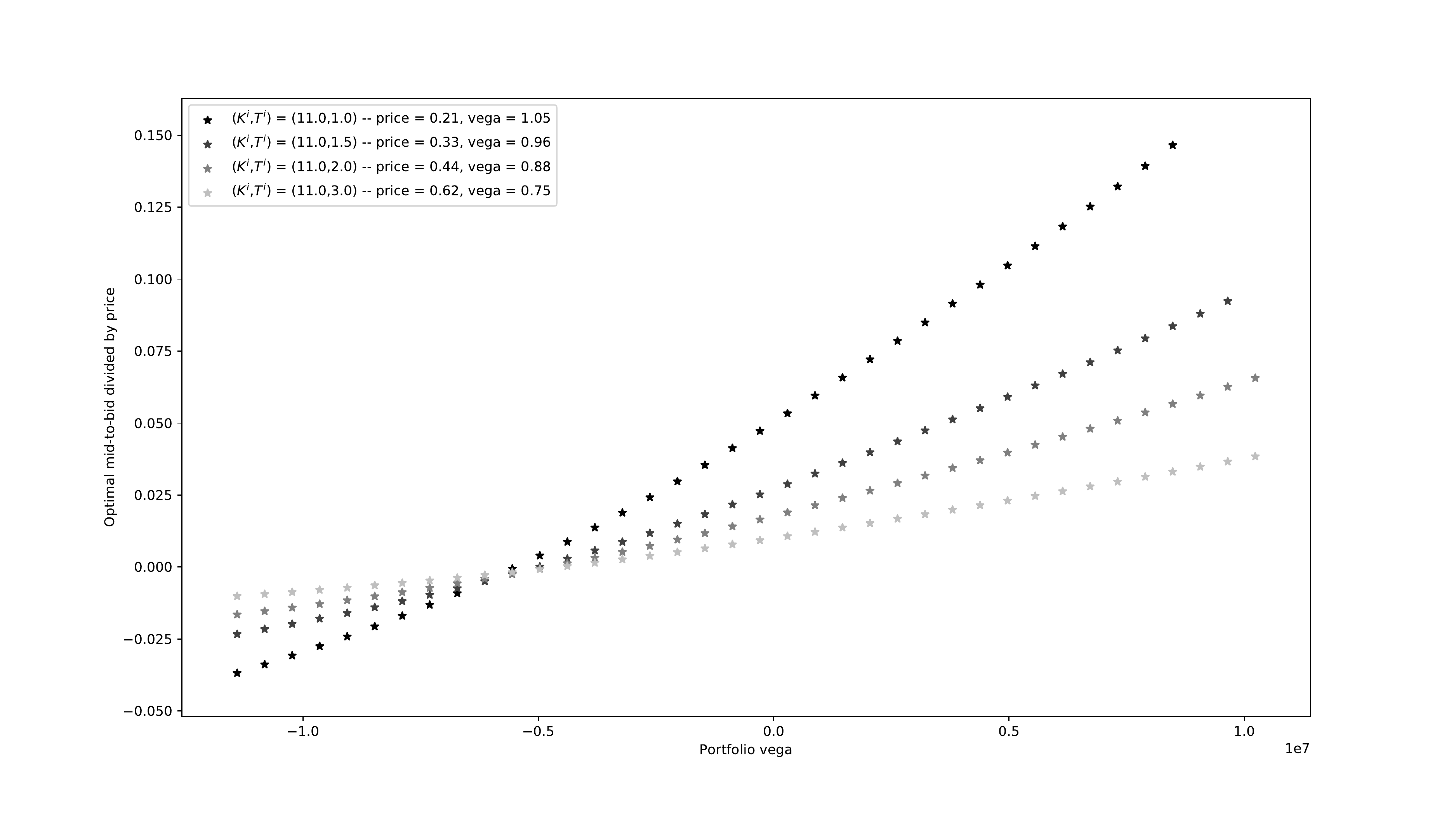}
\caption{Optimal mid-to-bid quotes divided by option price as a function of the portfolio vega for K=11.}
\label{optimal_quotes_11}
\end{center}
\end{figure}

\begin{figure}[!h]
\begin{center}
\includegraphics[width=0.92\textwidth]{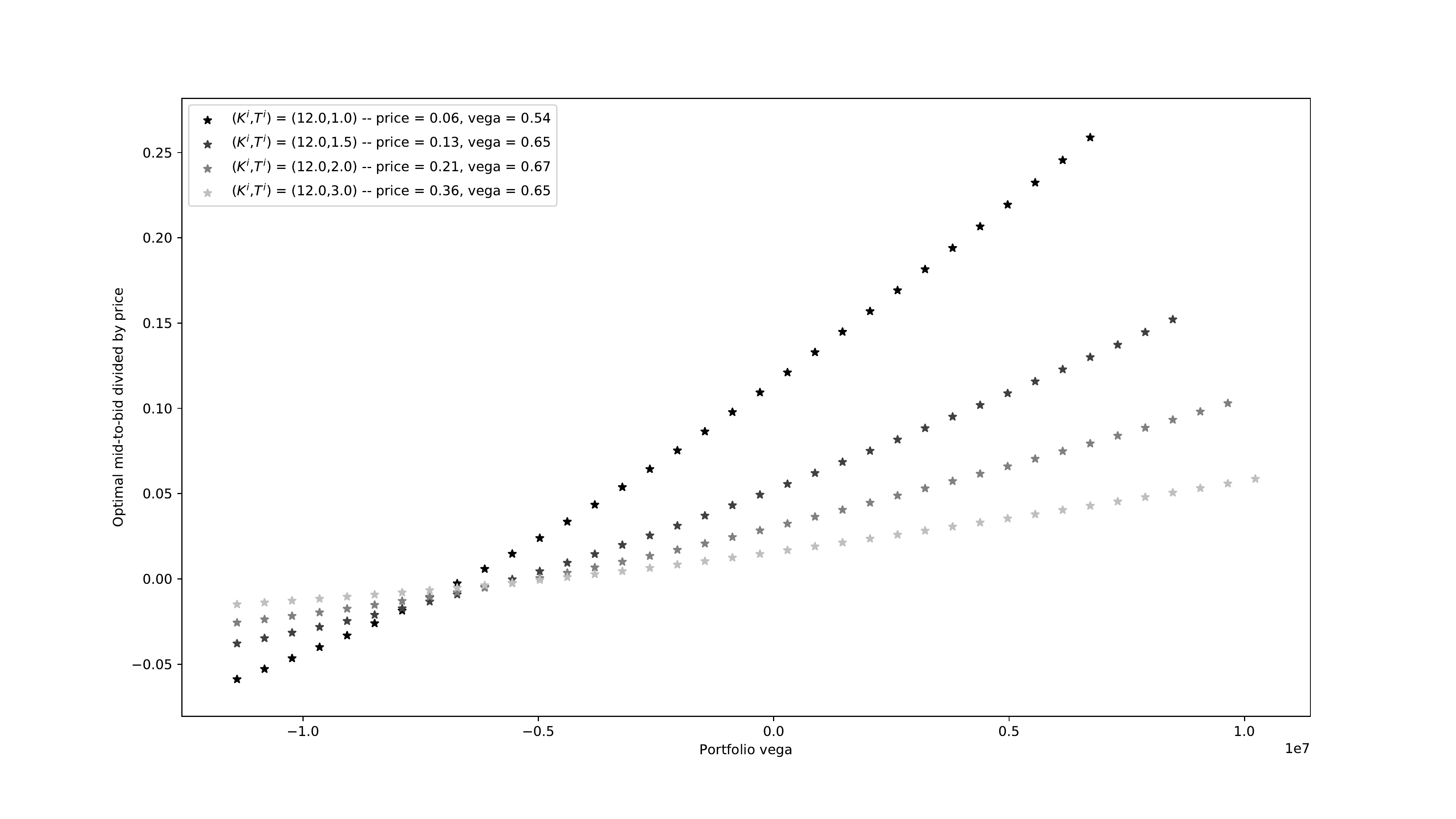}
\caption{Optimal mid-to-bid quotes divided by option price as a function of the portfolio vega for K=12.}
\label{optimal_quotes_12}
\end{center}
\end{figure}

Focusing on the asymptotic values, we now present in Figures \ref{optimal_quotes_8}, \ref{optimal_quotes_9}, \ref{optimal_quotes_10}, \ref{optimal_quotes_11}, and \ref{optimal_quotes_12}, the optimal mid-to-bid quotes as a function of the portfolio vega for each strike and maturity. More precisely, as the options we consider can have very different prices, we consider instead of the optimal bid quotes themselves the ratio between each optimal mid-to-bid quote and the price (at time $t=0$) of the corresponding option.

We clearly see that the optimal mid-to-bid quotes are increasing with the vega of the portfolio. This is expected since the options all have a positive vega: the higher the vega of the portfolio, the lower the willingness to buy another option, and therefore the lower the bid price (and the higher the mid-to-bid) proposed by the market maker.\\

\begin{figure}[!h]
\begin{center}
\includegraphics[width=0.92\textwidth]{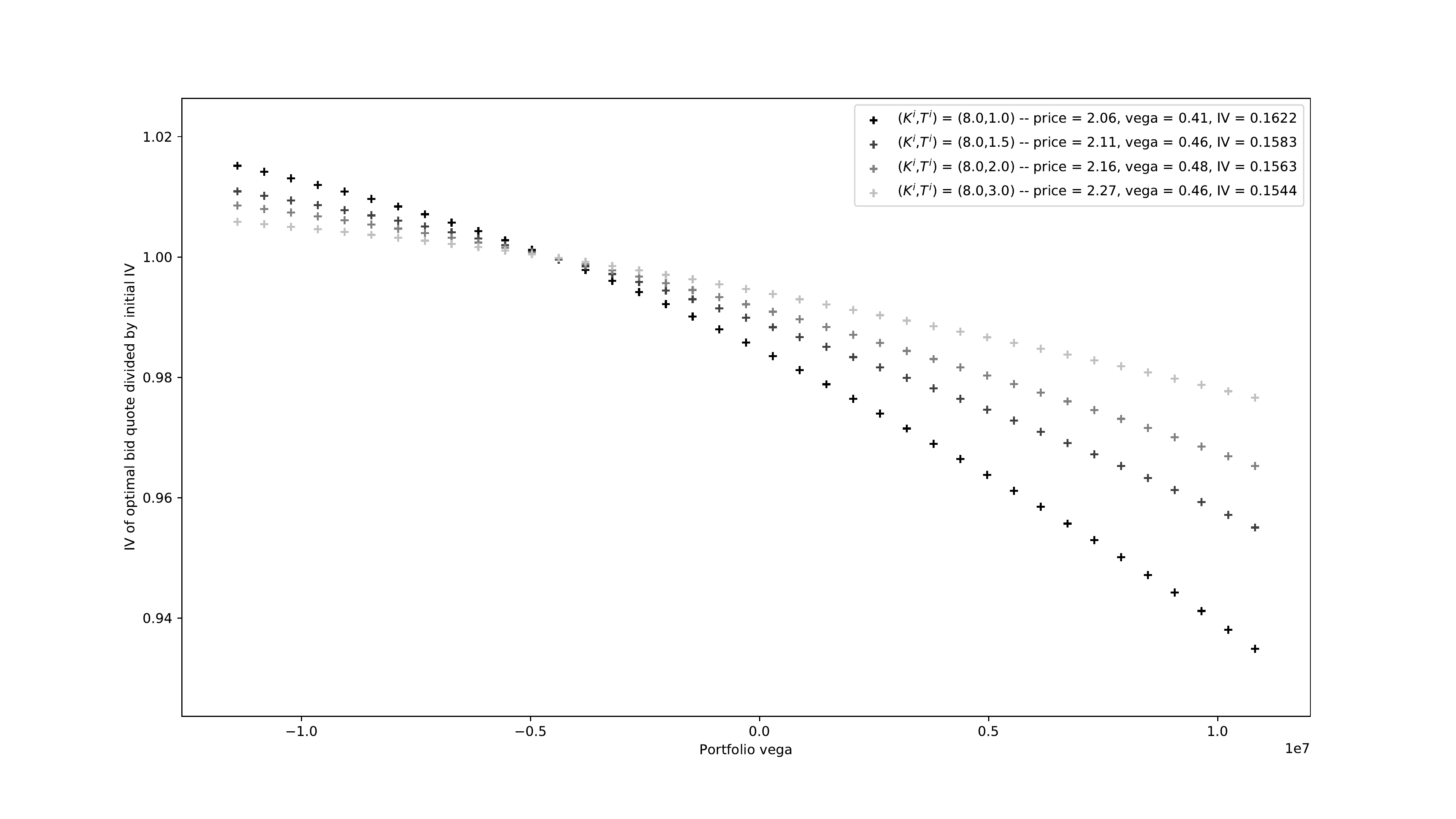}
\caption{Optimal (relative) bid implied volatility as a function of the portfolio vega for K=8.}
\label{opt_vol_8}
\end{center}
\end{figure}

\begin{figure}[!h]
\begin{center}
\includegraphics[width=0.92\textwidth]{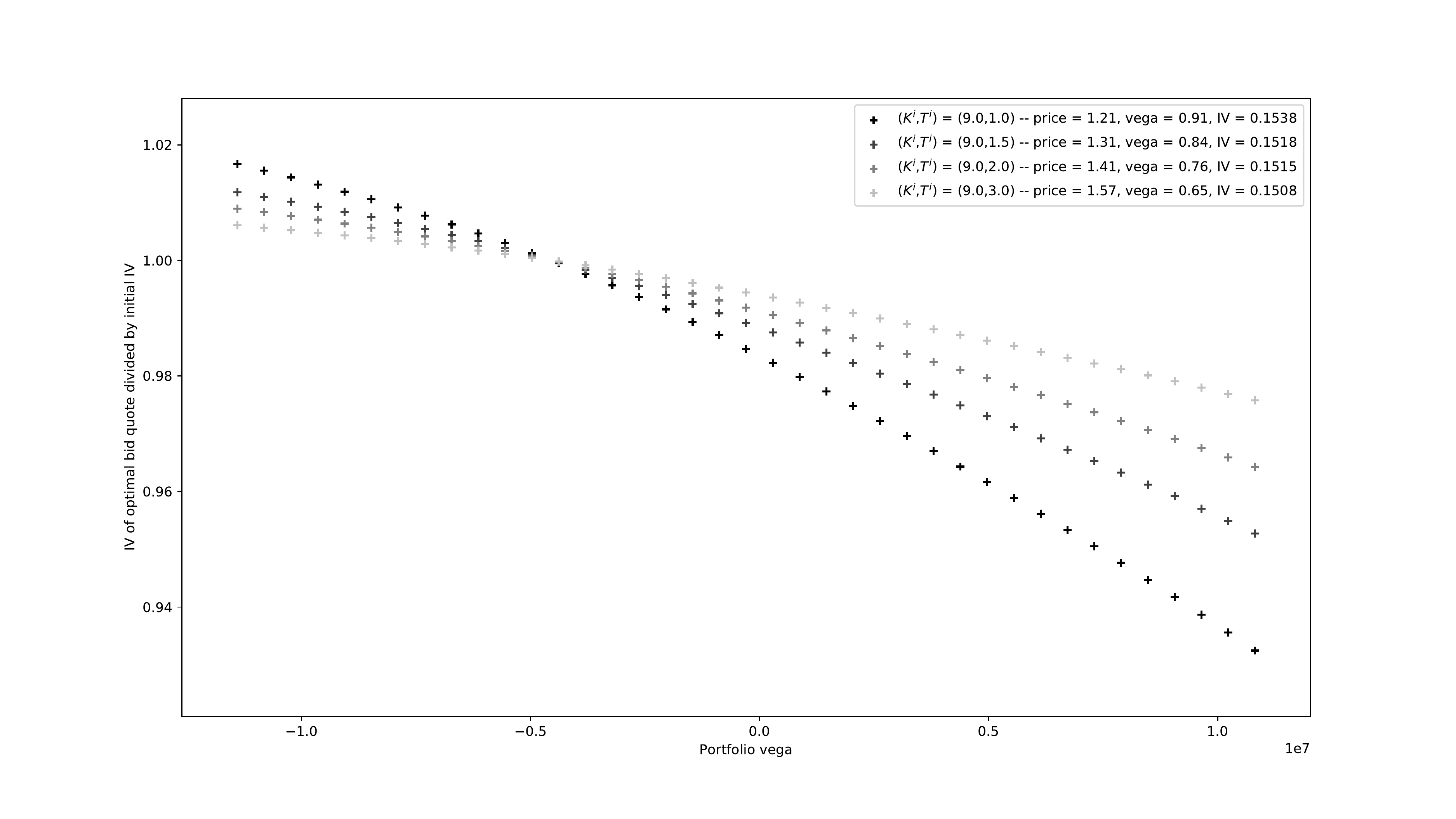}
\caption{Optimal (relative) bid implied volatility as a function of the portfolio vega for K=9.}
\label{opt_vol_9}
\end{center}
\vspace{2cm}
\end{figure}

\begin{figure}[!h]
\begin{center}
\includegraphics[width=0.92\textwidth]{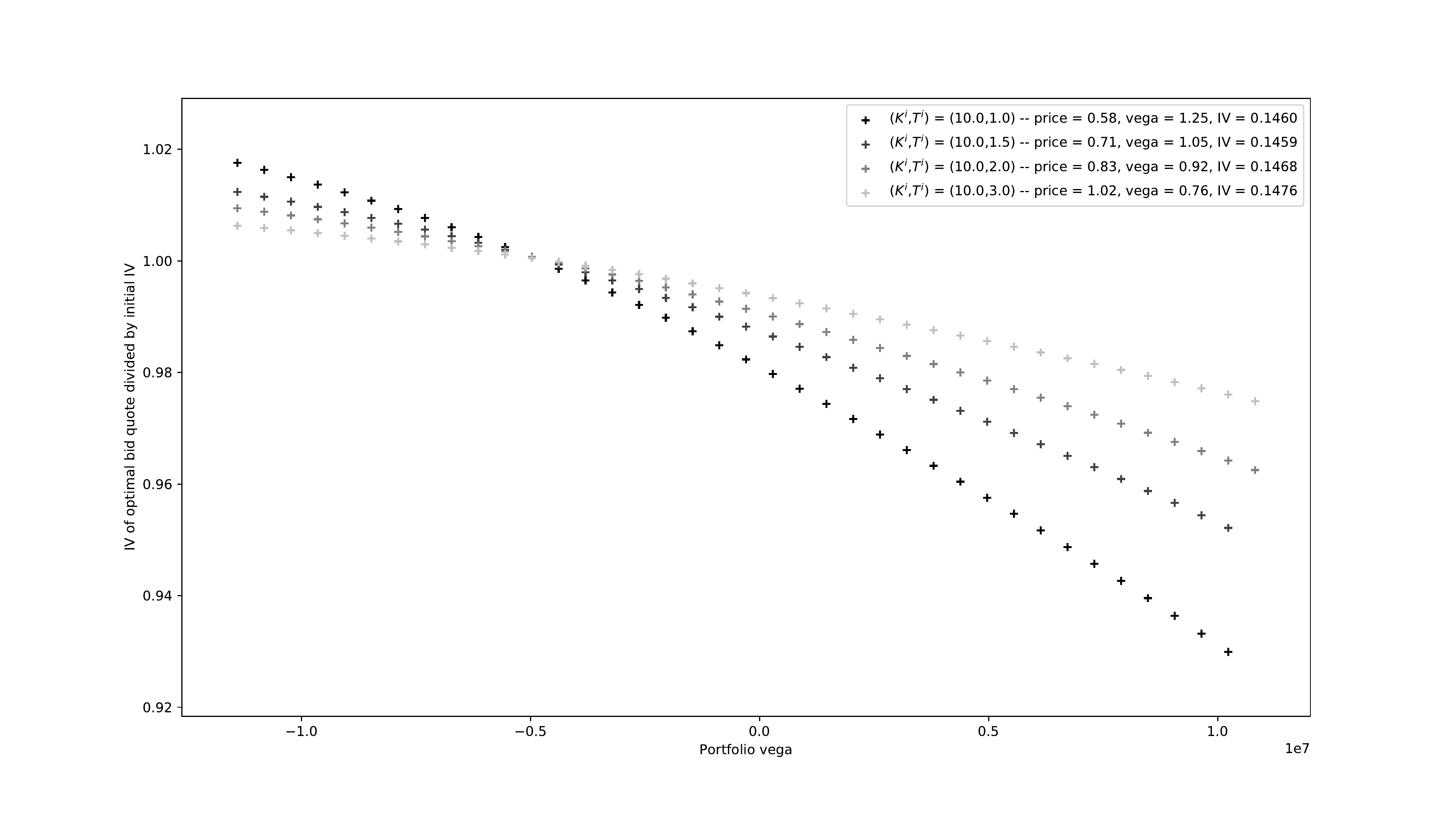}
\caption{Optimal (relative) bid implied volatility as a function of the portfolio vega for K=10.}
\label{opt_vol_10}
\end{center}
\end{figure}

\begin{figure}[!h]
\begin{center}
\includegraphics[width=0.92\textwidth]{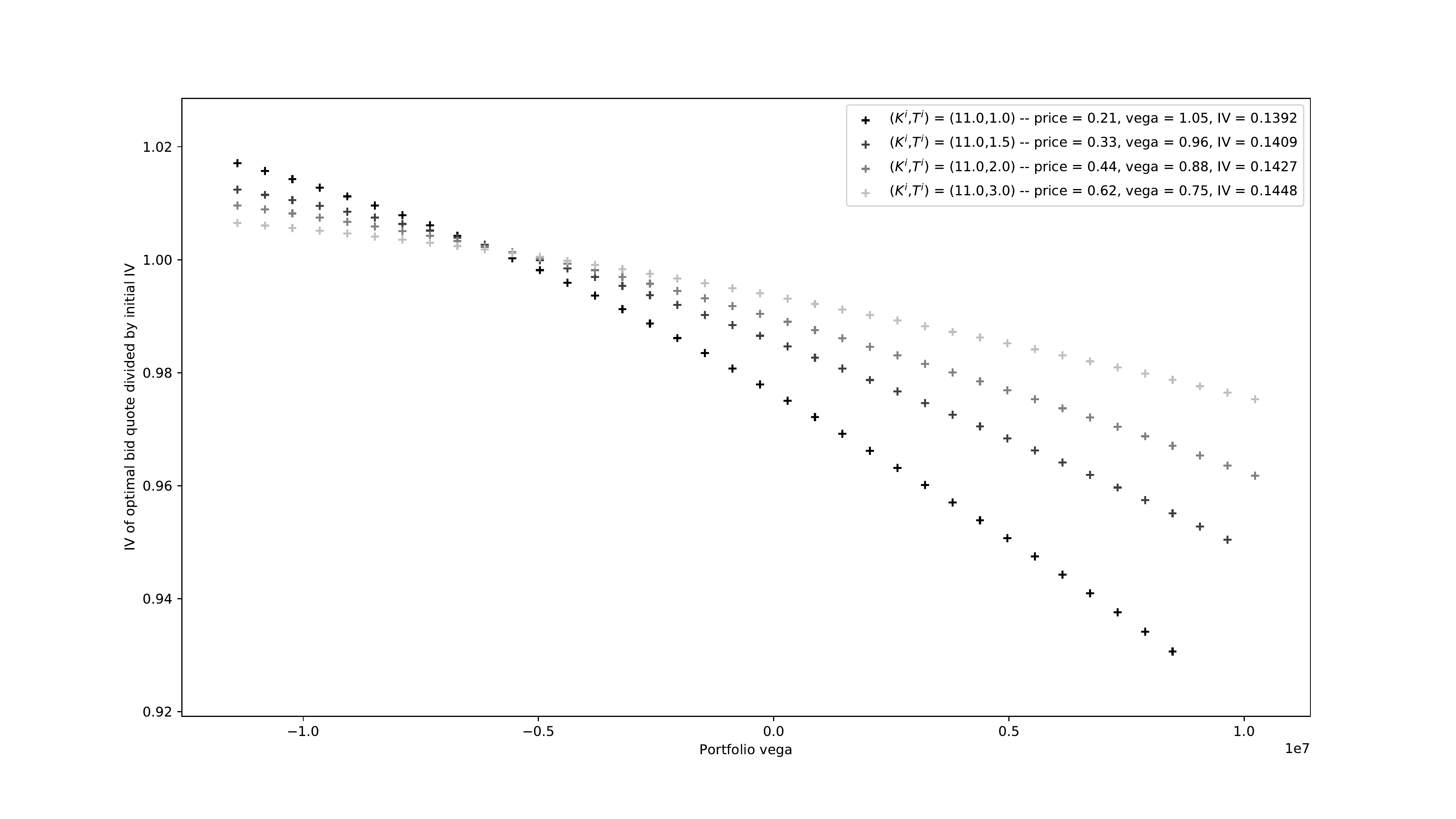}
\caption{Optimal (relative) bid implied volatility as a function of the portfolio vega for K=11.}
\label{opt_vol_11}
\end{center}
\vspace{2cm}
\end{figure}

\begin{figure}[!h]
\begin{center}
\includegraphics[width=0.92\textwidth]{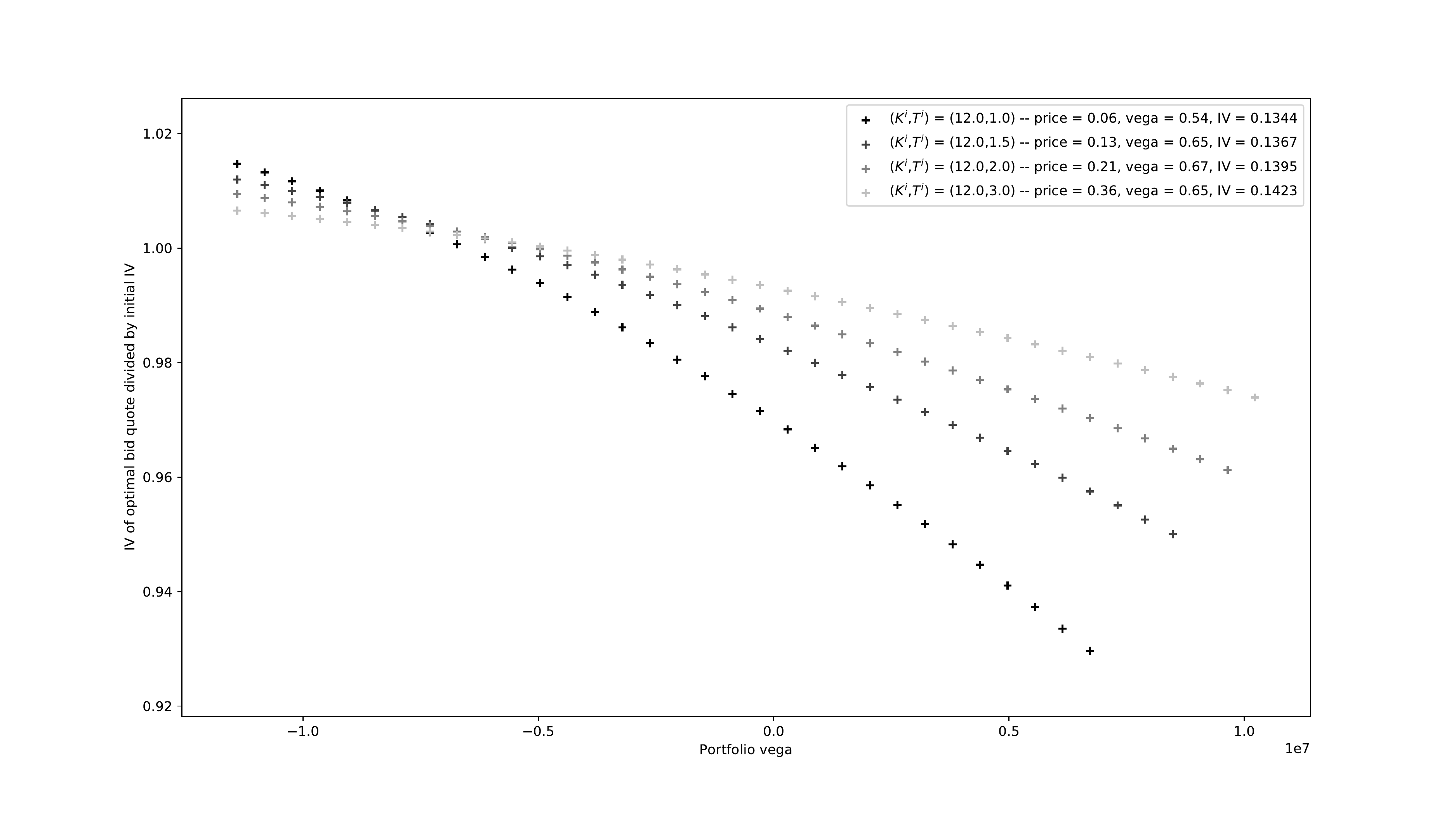}
\caption{Optimal (relative) bid implied volatility as a function of the portfolio vega for K=12.}
\label{opt_vol_12}
\end{center}
\end{figure}

Since option market makers usually reason in terms of implied volatility, we plot the same optimal bid quotes for the 20 options in terms of implied volatility (divided by the implied volatility at time $t=0$) in Figures \ref{opt_vol_8}, \ref{opt_vol_9}, \ref{opt_vol_10}, \ref{opt_vol_11}, and \ref{opt_vol_12}. We obtain of course decreasing curves.\\

\section*{Conclusion}

In this article, we tackled the problem of an option market maker dealing with options on a single underlying asset.\footnote{As noted when presenting this paper to a wide audience, our method could easily be extended to the case of multiple underlying assets using the same method as in \cite{bergault2019optimal} if the instantaneous variance processes of the different assets are driven by a few factors.} Using a constant-vega approximation, we showed how to reduce the problem to a low-dimensional functional equation whose solution can easily be approximated using an explicit Euler scheme and linear interpolation techniques. Furthermore, our method scales linearly in the number of options and can therefore be used with large books of options. Our method is illustrated by an example involving 20 European calls, but our model can be used with any European options.

\appendix

\section{Appendices}

\subsection{An alternative to the $\Delta$-hedging assumption}
\label{A1}
Throughout the body of this paper, we assumed that the market maker ensured $\Delta$-hedging. In this appendix, we show that this assumption can be relaxed without much change in the reasoning.\\

Let us introduce the process $(q_t^S)_{t\in[0,T]}$ representing the inventory of the market maker in the underlying asset. The dynamics of the cash process of the market maker $(X_t)_{t\in[0,T]}$ rewrites as
$$
 dX_{t}=\underset{i=1}{\overset{N}{\sum}}\bigg(\int_{\mathbb{R}_+^*}z\Big(\delta_{t}^{i,b}(z)N_{t}^{i,b}(dt,dz)+\delta_{t}^{i,a}(z)N_{t}^{i,a}(dt,dz)\Big)-\mathcal{O}_{t}^{i}dq_{t}^{i}\bigg)-S_t dq^S_{t}-d\big\langle q^S,S\big\rangle_t.
$$
The Mark-to-Market value of the portfolio writes
$$
V_{t}=X_{t}+q_t^S S_{t}+\underset{i=1}{\overset{N}{\sum}}q_{t}^{i}\mathcal{O}_{t}^{i}
$$
and its dynamics is
\begin{eqnarray*}
dV_{t}&=&\underset{i=1}{\overset{N}{\sum}}\left(\int_{\mathbb{R}_+^*}z\Big(\delta_{t}^{i,b}(z)N_{t}^{i,b}(dt,dz)+\delta_{t}^{i,a}(z)N_{t}^{i,a}(dt,dz)\Big)+ q_{t}^{i}\mathcal{V}_{t}^{i} \frac{a_{\mathbb{P}}(t,\nu_t) - a_{\mathbb{Q}}(t,\nu_t)}{2\sqrt{\nu_{t}}}dt + \frac{\xi}{2} q_{t}^{i}\mathcal{V}_{t}^{i}dW_{t}^{\nu}\right)\\
&&+\sqrt{\nu_t}S_t\left(\sum_{i=1}^N q_t^i \partial_S O^i(t,S_t,\nu_t) +q_t^S\right)dW_t^S.\\
\end{eqnarray*}
Denoting by $\Delta^\pi_t:=\sum_{i=1}^{N}q_t^{i}\partial_S O^i(t,S_t,\nu_t)$ the $\Delta$ of the market maker's portfolio at time $t$, our mean-variance optimization problem becomes
$$
\sup_{(\delta,q^S)\in \mathcal{A}'}\mathbb{E}\left[V_{T}\right]-\frac{\gamma}{2}\mathbb{V}\left[\int_0^T \frac{\xi}{2} \mathcal{V}_t^\pi dW_{t}^{\nu}+\sqrt{\nu_t}S_t\left(\Delta_t^\pi +q_t^S\right)dW_t^S \right],
$$
where
\begin{eqnarray*}
\mathcal{A}' = \left\{ (\delta_t,q^S_t)_{t \in [0,T]}\right. &|& \delta \text{ is a }\mathbb{F}\text{-predictable $\mathbb R^{2N}$-valued map bounded from below by } \delta_\infty\\
&& \left. \text{and }q^S \text{ is an } \mathbb{R}\text{-valued adapted process with } \mathbb{E}\left[\int_0^T \nu_tS^2_t\left(\Delta_t^\pi +q_t^S\right)^2 dt \right]< +\infty  \right\}.
\end{eqnarray*}

Noticing that
$$
    \mathbb{V}\!\left(\int_0^T  \frac{\xi}{2} \mathcal{V}_t^\pi dW_{t}^{\nu}\!+\!\sqrt{\nu_t}S_t\big(\Delta_t^\pi +q_t^S\big)dW_t^S\!\right)\!=\!\mathbb{E}\!\left[ \int_0^T \left(\frac{\xi^2}{4}{\mathcal{V}_t^\pi}^2 + \nu_t S_t^2\left(\Delta_t^\pi\!+\!q_t^S\right)^2\!+\!\rho \xi \mathcal{V}_t^\pi \sqrt{\nu_t}S_t\left(\Delta_t^\pi +q_t^S\right)\right)\mathrm{d}t \right],
$$
we easily see that the variance term is minimized for $q^S = {q^{S}}^*$ where
$$\forall t \in [0,T],\quad  {q^{S}_t}^* = - \Delta_t^\pi -\frac{\rho\xi\mathcal{V}_t^\pi}{2 \sqrt{\nu_t}S_t},$$
and that its minimum value is
$$(1-\rho^2)\int_0^T \frac{\xi^2}{4}{\mathcal{V}_t^\pi}^2 dt.$$
Therefore, the optimization problem boils down to
$$\sup_{\delta\in \mathcal{A}} \mathbb{E}\left[\int_{0}^{T}\left(\left(\underset{i=1}{\overset{N}{\sum}}\underset{j=a,b}{\sum}\int_{\mathbb{R}_+^*}z\delta_{t}^{i,j}(z)\Lambda^{i,j}(\delta_{t}^{i,j}(z))\mathds{1}_{\left\{|\mathcal{V}^{\pi}_t-\psi(j)z\mathcal{V}^{i}|\leq \overline{\mathcal V}\right\}}\mu^{i,j}(dz)\right)\right.\right.$$
$$\left.\left. \vphantom{\sup_{\delta\in \mathcal{A}} \mathbb{E}\left[\int_{0}^{T}\left(\left(\underset{i=1}{\overset{N}{\sum}}\underset{j=a,b}{\sum}\int_{\mathbb{R}_+^*}z\delta_{t}^{i,j}(z)\Lambda^{i,j}(\delta_{t}^{i,j}(z))\mathds{1}_{\left\{|\mathcal{V}^{\pi}_t-\psi(j)z\mathcal{V}^{i}|\leq \overline{\mathcal V}\right\}}\mu^{i,j}(dz)\right)\right.\right.} + \mathcal{V}^{\pi}_t \frac{a_{\mathbb{P}}(t,\nu_t) - a_{\mathbb{Q}}(t,\nu_t)}{2\sqrt{\nu_{t}}} -\frac{\gamma\xi^{2}}{8}(1-\rho^2) {\mathcal{V}^\pi_t}^2\right) dt\right],
$$
and we recover the same optimization problem as in the body of the paper, except that the volatility of volatility parameter $\xi$ is multiplied by $\sqrt {1-\rho^2}$ to account for the reduction of risk made possible by the optimal trading strategy in the underlying asset in presence of vol-spot correlation.

\subsection{Beyond the constant-vega assumption}
\label{A2}
In this appendix we propose a method to relax our main assumption: the constant-vega approximation.\\

If, for all $i \in \{1,\ldots,N\}$, the process $\left( \mathcal{V}^i_t \right)_{t \in [0,T]}$ stays close to its initial value $\mathcal{V}^i_0 =: \mathcal{V}^i$, then it is reasonable to consider a perturbative approach around the constant-vega approximation. In particular, instead of assuming that $\sum_{i=1}^N q^i_t \partial_{\sqrt{\nu}}O^{i}(t,S_{t},\nu_{t}) \simeq \sum_{i=1}^N q^i_t \partial_{\sqrt{\nu}}O^{i}(0,S_{0},\nu_{0}) = \sum_{i=1}^N q^i_t \mathcal{V}^i  = \mathcal{V}_t^\pi$, we consider the expansion $$\sum_{i=1}^N q^i_t \partial_{\sqrt{\nu}}O^{i}(t,S_{t},\nu_{t}) = \mathcal{V}_t^\pi + \varepsilon \mathcal{W}(t,S_{t},\nu_{t}, q_t)$$
and we consider an expansion of the value function $u$ of the following form:
$$u(t,S,\nu,q) = v\left(t,\nu, \sum_{i=1}^N q^i \partial_{\sqrt{\nu}}O^{i}(0,S_{0},\nu_{0})\right) + \varepsilon \varphi(t,S,\nu,q) = v\left(t,\nu, \sum_{i=1}^N q^i \mathcal{V}^i\right) + \varepsilon \varphi(t,S,\nu,q).$$

Assuming that $\mathcal{Q} = \mathbb{R}^N$ and noting that the Hamilton-Jacobi-Bellman equation associated with $u$ is
\begin{align*}
0 =\ & \partial_t u (t,S,\nu,q) + a_{\mathbb{P}}(t,\nu) \partial_{\nu}u(t,S,\nu,q) + \frac 12 \nu S^2 \partial^2_{SS}u(t,S,\nu,q) + \frac 12 \nu \xi^2 \partial^2_{\nu \nu}u(t,S,\nu,q) + \rho \nu S \xi \partial^2_{\nu S}u(t,S,\nu,q)\\
& + \frac{a_{\mathbb{P}}(t,\nu) - a_{\mathbb{Q}}(t,\nu)}{2\sqrt{\nu}} \sum_{i=1}^N q^i \partial_{\sqrt{\nu}}O^{i}(t,S,\nu) - \frac{\gamma  \xi^2}{8} \left( \sum_{i=1}^N q^i \partial_{\sqrt{\nu}}O^{i}(t,S,\nu) \right)^2 \\
& + \sum_{i=1}^N \sum_{j=a,b} \int_{\mathbb{R}_+^*}z H^{i,j} \left(\frac{u(t,S,\nu,q) - u(t,S,\nu,q-\psi(j)z e^i)}{z} \right)\mu^{i,j}(dz),
\end{align*}
with terminal condition equal to $0$, the first-order term in $\varepsilon$ in the Taylor expansion gives
\begin{align*}
0  =\ & \partial_t \varphi(t,S,\nu,q) +a_{\mathbb{P}}(t,\nu) \partial_{\nu}\varphi(t,S,\nu,q) + \frac 12 \nu S^2 \partial^2_{SS}\varphi(t,S,\nu,q) + \frac 12 \nu \xi^2 \partial^2_{\nu \nu} \varphi(t,S,\nu,q) + \rho \nu S \xi \partial^2_{\nu S}\varphi(t,s,\nu,q)\\\
+& \frac{a_{\mathbb{P}}(t,\nu)-a_{\mathbb{Q}}(t,\nu)}{2\sqrt{\nu}} \mathcal{W}(t,S,\nu,q)  - \frac{\gamma \xi^2}{4} \mathcal{W}(t,S,\nu,q) \sum_{i=1}^N q^i \mathcal{V}^i  \\\
+&\sum_{i=1}^N \sum_{j=a,b}\int_{\mathbb{R}_+^*} {H^{i,j}}'\left(\frac{v(t,\nu,\sum_{l=1}^N q^l \mathcal{V}^l)-v(t,\nu,\sum_{l=1}^N q^l \mathcal{V}^l - \psi(j)z \mathcal{V}^i )}{z}\right)\ \\
& \qquad \qquad \qquad \times \left( \varphi(t,S,\nu,q)-\varphi(t,S,\nu,q-\psi(j)ze^i)\right) \mu^{i,j}(dz),
\end{align*}
with terminal condition equal to $0$.\\

This equation is linear and therefore $\varphi(t,S,\nu,q)$ admits a Feynman-Kac representation that tames the curse of dimensionality for practical applications:
$$\varphi(t,S,\nu,q) = \mathbb{E}_{(t,S,\nu,q)} \left[ \int_t^T \left(\frac{a_{\mathbb{P}}(s,\nu_s)-a_{\mathbb{Q}}(s,\nu_s)}{2\sqrt{\nu_s}} \mathcal{W}(s,S_s,\nu_s,q_s)  - \frac{\gamma \xi^2}{4} \mathcal{W}(s,S_s,\nu_s,q_s) \sum_{i=1}^N q_s^i \mathcal{V}^i \right)  ds \right]$$
where, for each $i \in \{1,\ldots,N\},$ the processes $N^{i,b}$ and $N^{i,a}$ have respective intensities
\begin{align*}
    \tilde{\lambda}^{i,b}_t(dz) &= -{H^{i,b}}' \left( \frac{v(t,\nu_t,\sum_{l=1}^N q^l_{t-} \mathcal{V}^l)-v(t,\nu_t,\sum_{l=1}^N q^l_{t-} \mathcal{V}^l +z \mathcal{V}^i)}{z} \right) \mu^{i,b}(dz),\\
    \tilde{\lambda}^{i,a}_t(dz) &= -{H^{i,a}}' \left( \frac{v(t,\nu_t,\sum_{l=1}^N q^l_{t-} \mathcal{V}^l) - v(t,\nu_t,\sum_{l=1}^N q^l_{t-} \mathcal{V}^l - z \mathcal{V}^i)}{z} \right) \mu^{i,a}(dz),
\end{align*}
with as before $dq^i_t = \int_{\mathbb{R}_+^*}z\left(N^{i,b}(dt,dz) - N^{i,a}(dt,dz)\right)$.\\

Subsequently, the function $\varphi$ can be computed using a Monte-Carlo algorithm and quotes accounting for the variation of the vegas can therefore be computed (to the first order in $\varepsilon$).

\subsection{On the construction of the processes $N^{i,b}$ and $N^{i,a}$}
\label{A3}
Let us consider a new filtered probability space $\big(\Omega,\mathcal{F},(\mathcal{F}_t)_{t\in \mathbb{R}_{+}},\tilde{\mathbb{P}}\big)$. For the sake of simplicity, assume that there is only one option (the generalization is straightforward). Let us introduce $\bar{N}^b$ and $\bar{N}^a$ two independent compound Poisson processes of intensity 1 whose increments follow respectively the distributions $\mu^b(dz)$ and $\mu^a(dz)$ with support on $\mathbb{R}_+^*$. We denote by $\bar{N}^b(dt,dz)$ and $\bar{N}^a(dt,dz)$ the associated random measures.  Let $N^b$ and $N^a$ be two processes, starting at 0, solutions of the coupled stochastic differential equation:
\begin{align*}
    dN^b_t = \int_{\mathbb{R}_+^*}\mathds{1}_{\left\{N^b_{t-} - N^a_{t-} + z \in \mathcal{Q}\right\}} \bar{N}^b(dt,dz),\\
    dN^a_t = \int_{\mathbb{R}_+^*} \mathds{1}_{\left\{N^b_{t-} - N^a_{t-} - z \in \mathcal{Q}\right\}} \bar{N}^a(dt,dz).
\end{align*}
Then, under $\tilde{\mathbb{P}},$ $N^b$ and $N^a$ are two $\mathbb{R}^*_+$-marked point processes with respective intensity kernels $$\lambda^b_t(dz) = \mathds{1}_{\left\{q_{t-} + z \in \mathcal{Q}\right\}} \mu^b(dz) \quad \text{and} \quad \lambda^a_t(dz) =  \mathds{1}_{\left\{q_{t-} - z \in \mathcal{Q}\right\}}\mu^a(dz),$$
where $q_t := N^b_t - N^a_t.$ We denote by $N^b(dt,dz)$ and $N^a(dt,dz)$ the associated random measures. For each $\delta \in \mathcal{A}$, we introduce the probability measure $\tilde{\mathbb{P}}^\delta$ given by the Radon-Nikodym derivative
\begin{equation}
\frac{d\tilde{\mathbb{P}}^\delta}{d\tilde{\mathbb{P}}} \Big|_{\mathcal{F}_t} = L_t^{\delta},
\end{equation}
where $\left(L_t^{\delta} \right)_{t\geq 0}$ is the unique solution of the stochastic differential equation
\begin{equation}
    dL_t^{\delta} = L_{t-}^{\delta} \left( \int_{\mathbb{R}_+^*} \left(\Lambda^b(\delta^b(t,z))-1 \right)\tilde{N}^b(dt,dz) + \int_{\mathbb{R}_+^*} \left(\Lambda^a(\delta^a(t,z))-1 \right)\tilde{N}^a(dt,dz) \right),\nonumber
\end{equation}
with $L_0^{\delta} = 1$, where $\tilde{N}^b$ and $\tilde{N}^a$ are the compensated processes associated with $N^b$ and $N^a$ respectively.\\

We then know from Girsanov theorem that under $\tilde{\mathbb{P}}^\delta$, the jump processes $N^{b}$ and $N^{a}$ have respective intensity kernels
$$\lambda^{\delta,b}_t(dz)  = \Lambda^b(\delta^b(t,z)) \mathds{1}_{\left\{q_{t-} + z\in \mathcal{Q}\right\}} \mu^b(dz) \quad \text{and} \quad \lambda^{\delta,a}_t(dz)  = \Lambda^a(\delta^a(t,z)) \mathds{1}_{\left\{q_{t-} - z \in \mathcal{Q}\right\}} \mu^a(dz)$$ as in the body of the paper.

\end{document}